\documentclass[prx,twocolumn,superscriptaddress]{revtex4-1}
\usepackage[usenames,dvipsnames,svgnames,table]{xcolor}
\usepackage{hyperref}
\usepackage{amsmath, amsfonts,amssymb, mathtools} 
\hypersetup{colorlinks=true,  linkcolor=NavyBlue, citecolor=PineGreen,urlcolor=cyan}
\usepackage{physics}
\usepackage{graphicx}

\usepackage[mathscr]{euscript}

\usepackage{bm}

\newcommand{\bracket}[1]{\left\langle #1\right\rangle}
\newcommand{\beeq}[1] {\begin{equation}\begin{split}#1\end{split}\end{equation}}

\begin{document}

\title{Non-convex image reconstruction via Expectation Propagation}

\author{Anna Paola Muntoni}
\affiliation{Department of Applied Science and Technologies (DISAT), Politecnico di Torino, Corso Duca Degli Abruzzi 24, Torino, Italy}
\affiliation{Laboratoire de physique th\'eorique, D\'epartement de physique de l'ENS, \'Ecole normale sup\'erieure, PSL University, Sorbonne Universit\'e, CNRS, 75005 Paris, France}
\author{Rafael D\'iaz Hern\'andez Rojas} 
\affiliation{Dipartimento di Fisica, Sapienza University of Rome, P.le Aldo Moro 5, I-00185 Rome, Italy}
\author{Alfredo Braunstein}
\thanks{Joint last authors.}
\affiliation{Department of Applied Science and Technologies (DISAT), Politecnico di Torino, Corso Duca Degli Abruzzi 24, Torino, Italy}
\affiliation{Italian Institute for Genetic Medicine (form. HuGeF), Via Nizza 52, Torino, Italy}
\affiliation{INFN Sezione di Torino, Via P. Giuria 1, I-10125 Torino, Italy}
\affiliation{Collegio Carlo Alberto, Via Real Collegio 30, Moncalieri, Italy}

\author{Andrea Pagnani}
\thanks{Joint last authors.}
\affiliation{Department of Applied Science and Technologies (DISAT), Politecnico di Torino, Corso Duca Degli Abruzzi 24, Torino, Italy}
\affiliation{Italian Institute for Genetic Medicine (form. HuGeF), Via Nizza 52, Torino, Italy}
\affiliation{INFN Sezione di Torino, Via P. Giuria 1, I-10125 Torino, Italy}
\author{Isaac P\'erez Castillo\footnote[x]{Joint last authors}}
\thanks{Joint last authors.}
\affiliation{Departamento de F\'isica Cu\'antica y F\'otonica, Instituto de F\'isica, UNAM, P.O. Box 20-364, 01000 Cd. Mx., M\'exico}
\affiliation{London Mathematical Laboratory, 8 Margravine Gardens, W6 8RH London, United Kingdom}

\begin{abstract}
  Tomographic image reconstruction can be mapped to a  problem of finding solutions to a large system of linear equations which maximize a function that includes \textit{a priori} knowledge regarding features of typical images such as smoothness or sharpness. This maximization can be performed with standard local optimization tools when the function is concave, but it is generally intractable for realistic priors, which are non-concave. We introduce a new method to reconstruct images  obtained from Radon projections by using Expectation Propagation,  which allows us to reframe the problem from an Bayesian inference perspective. We show, by means of extensive simulations, that, compared to state-of-the-art algorithms for this task, Expectation Propagation paired with very simple but non log-concave priors, is often able to reconstruct images up to a smaller error while using a lower amount of information per pixel. We provide estimates for the critical rate of information per pixel above which recovery is error-free by means of simulations on ensembles of phantom and real images.
\end{abstract}

\maketitle

\section{Introduction}
The classical problem in image reconstruction consists in recovering the density of an object's 2D slice from a set of Radon transformations. These correspond to a set of projections on a plane, which can be usually identified with the absorption of radiation by the object along a given line  of response. In the ideal case, and from a purely mathematical point of view, the original image can be reconstructed when enough projections are available by applying the inverse Radon transform, a method usually referred to as filtered back-projection \cite{Avinash1988}. However, in real scenarios, detectors and radiation sources have an actual finite size, data collection is restricted to a short time window and measurements are naturally noisy. With limited and/or noisy information, it is only possible, in principle, to reconstruct a finite resolution discretization $\bm{x}=(x_1,\ldots, x_N)^T\in\mathbb{R}^N$ of the image from a finite set of Radon projections. In these more realistic scenarios, Algebraic Reconstruction Techniques (ART) are normally used. Under reasonable working hypotheses, discretized images that are compatible with the measurements are the ones satisfying a set of linear relations:
\beeq{
\bm{A} \bm{x}=\bm{p}\,, \label{eq:reconstruction_as_a_linear_problem}
}
where $\bm{p}=(p_1,\ldots,p_M)^T\in\mathbb{R}^M$ is the $M$-dimensional measurements vector of projection data, and $\bm{A}=(a_{ij})\in\mathbb{R}^{M\times N}$ is the so-called \textit{projection matrix}. Here, variable $x_j$ represents the density of the image at the position of pixel $j$, while the entries $a_{ij}$ of the matrix $\bm{A}$ correspond to the length of the intersection of the $i$-th projection ray with the $j$-th pixel or, in other terms, to the contribution of the $j$-th pixel to the total attenuation along the $i$-th ray. In the field of image reconstruction, a pivotal role is played by algorithms capable of providing accurate reconstructions with the lowest possible number of measurements $M$. Often in medical imaging, one faces practical constraints posed by the acquisition system, and/or the need to mitigate the dangerous effects of ionizing radiation exposure. In the following we will mostly concentrate on this {\em limited data} regime that corresponds to an under-determined system of equations in Eq. \eqref{eq:reconstruction_as_a_linear_problem}, \textit{i.e.} $M<N$, where the system has infinitely many solutions in the noiseless regime and none in the noisy one. Using the ART algorithm \cite{Kaczmarz1937,Gordon1970}, one can, however, obtain an approximate reconstruction of the image $\bm{x}$ by iteratively minimizing the $\ell_2$ error, $||\bm{A}\bm{x}-\bm{p}||_2$. Its performance can be drastically enhanced by combining it with the Total Variation (TV) method \cite{Sidky2006} which relies on our \textit{a priori} knowledge that realistic images have intrinsic structure, in particular smoothness, that can be encoded by means of a $\ell_1$ sparsity regularization on the (discrete) gradient of the image.

The reconstruction problem can be alternatively recast in the language of Bayesian inference by considering the posterior probability distribution of images for a given vector of measurements $\bm{p}$
\begin{equation}
  P(\bm{x}|\bm{p})=P(\bm{p})^{-1} P(\bm{p}|\bm{x}) P_0(\bm{x})\, .
  \label{eq:general_posterior}
\end{equation}
The reconstructed image is often given by the \textit{Maximum a Posteriori} (MAP) estimation  $\bm{x}^\star=\text{arg max}_{x}P(\bm{x}|\bm{p})$. The \emph{likelihood} term $P(\bm{p}|\bm{x})$ corresponds to the discretized model  in Eq. \eqref{eq:reconstruction_as_a_linear_problem}  and takes the form  $P(\bm{p}|\bm{x}) = \delta(\bm{A}\bm{x}-\bm{p})$ for the noiseless case \cite{Gouillart2013}. The \emph{prior} $P_0(\bm{x})$ plays a crucial role as it allows to include further information complementing the set of measurements, making the reconstruction possible in the under-determined regime. Fairly intuitively, a smaller amount of image-specific information is needed to perform the reconstruction provided we have access to a more informative prior on the class of images. Both $\ell_2$ and $\ell_1$ regularizations can be mapped in this framework as log-concave priors that admit the computation of the corresponding MAP estimates by means of standard convex optimization techniques. The mapping between Eq. \eqref{eq:general_posterior} and $\ell_p$ regularization for $p>0$ is straightforward by considering $P_0(\bm{x})\propto \exp(-\lambda \|\bm{x}\|_p)$, as the measure then concentrates on the minima of $\ell_p$ for $\lambda\to\infty$. If measurements are affected by independent additive Gaussian noise (although other noise models can be assumed), the likelihood reads  $P(\bm{p}|\bm{x})\propto\exp[-\frac{\beta}{2}(\bm{p}-\bm{A}\bm{x})^2]$, where ${\beta}$ is the inverse variance of the noise distribution.  Note that in general, the mean value of $\bm{x}$ of the posterior distribution  $ P(\bm{x}|\bm{p})$ is the vector that minimizes the mean square error and should be generally preferred to the MAP estimation.

A thorough analysis of actual tomographic and natural images reveals that in many cases the statistics of pixel intensities are ill-fitted by trivial log-concave functions \cite{Bouman1993, FreemanJones2002, Sardy2004, Tanaka2012} but can, in principle, be well-fitted by priors involving non log-concave terms. This renders apparently the MAP estimate a computationally formidable task as it leads to a non-convex optimization problem. 

Recently, ground-breaking applications of statistical mechanics techniques to non-convex computational problems have yielded very efficient algorithms and reliable methods to make the computation of marginals of complicated multivariate distributions computationally tractable. These techniques have been recently and successfully applied to image reconstruction  of binary images for the case of discrete tomography \cite{Gouillart2013}. Here it was noted that the Belief Propagation (BP) algorithm provides better reconstruction than TV in some cases, especially in the high noise regime. Statistical techniques such as BP have the additional advantage of being able to deal more efficiently in the imperfect reconstruction regime than optimization methods, since the maximum probability point may be uninformative when the posterior distribution is not very concentrated.  For example, the image that minimizes the average quadratic error is given by the posterior pixels averages, i.e. the first moments of the marginal posterior distributions. The BP algorithm, however, relies on the Bethe-Peierls approximation which is inaccurate in many realistic scenarios. Although corrections to the Bethe-Peierls approximation abound in the literature \cite{Rizzo2005,Pelizzola2005,Weinwright2005, Parisi2006}, they become, more often than not, impractical for current applications. Our main purpose is to introduce a family of priors with a corresponding family of algorithms based on Expectation Propagation (EP), whose reconstruction performance surpasses the ones obtained with standard log-concave priors and standard local optimization algorithms. EP, originally introduced in \cite{Opper2001,Minka2001,Opper2004,Heskes2005} gives additionally a natural probabilistic framework to maximize the inference performance in the imperfect reconstruction regime. In particular, it allows to compute an approximation of the posterior marginal distribution and the posterior average, allowing in principle for  a  more accurate reconstruction. It should be noted that $\ell_p$ regularization with $0<p<1$ for tomographic reconstruction has been considered at least in \cite{sidky_image_2007}. However, in this latter work, the proposed reconstruction algorithm is based on a local optimizer and it lacks the probabilistic framework and interpretation proposed here.

This work is organised as follows: in Sect. \ref{sec:EP} we introduce the method of EP together with the different priors we have used when reconstructing images. Sect. \ref{sec:parameters} is dedicated to explain how one can estimate the various parameters of the method based solely on probabilistic arguments. Results of our approach for phantom and real images are presented and discussed in Sect. \ref{sec:results}. We end up with some concluding remarks in Sect. \ref{sec:discussion}. Thorough mathematical derivations, together with details of previous reconstructing algorithms, can be found in the appendices.

\section{Implementation of Prior knowledge and the method of Expectation Propagation}
\label{sec:EP}
Before discussing the method of EP to approximate the posterior distribution $P(\bm{x}|\bm{p})$, it is crucial to have a prior distribution  $P_0(\bm{x})$ that captures reasonably well some of the typical properties of the images we aim to reconstruct. To achieve this, we assume that the prior $P_0(\bm{x})$ can be written as a product $P_0(\bm{x}) \propto P_0^{(\text{single})}(\bm{x})P_0^{(\text{pair})}(\bm{x})$, where the factor $P_0^{(\text{single})}(\bm{x})$ imposes independent local constraints designed to capture the concrete nature and support of the pixels involved, whereas   $P_0^{(\text{pair})}(\bm{x})$ contains all priors that can be written as product of probability distribution over pairs of variables and it is supposed to model the highly correlated nature among pixels in real images.

\begin{figure}
  \begin{center}
    \includegraphics[width=4cm, height=3.4cm]{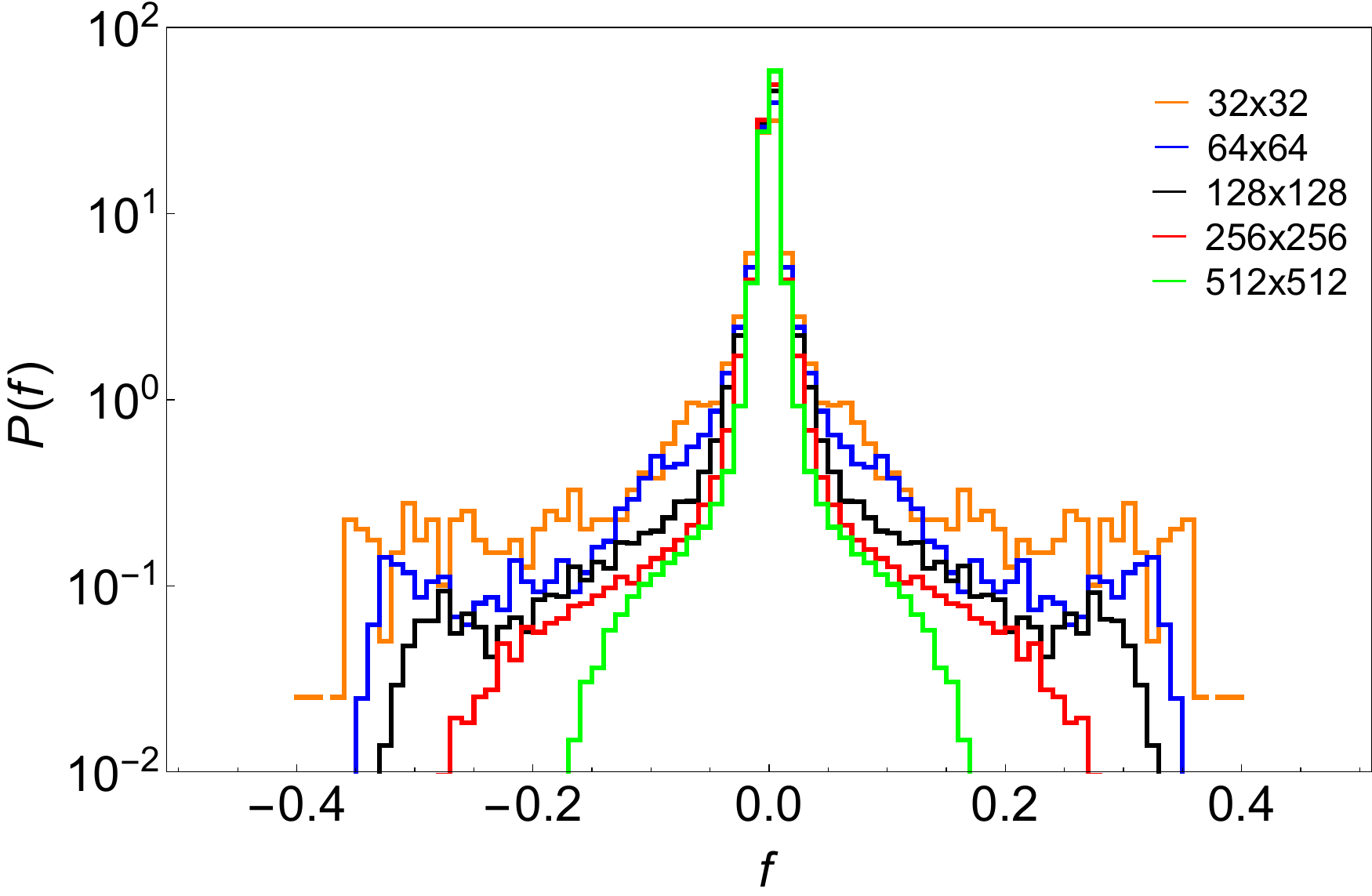}\includegraphics[width=4cm, height=3.4cm]{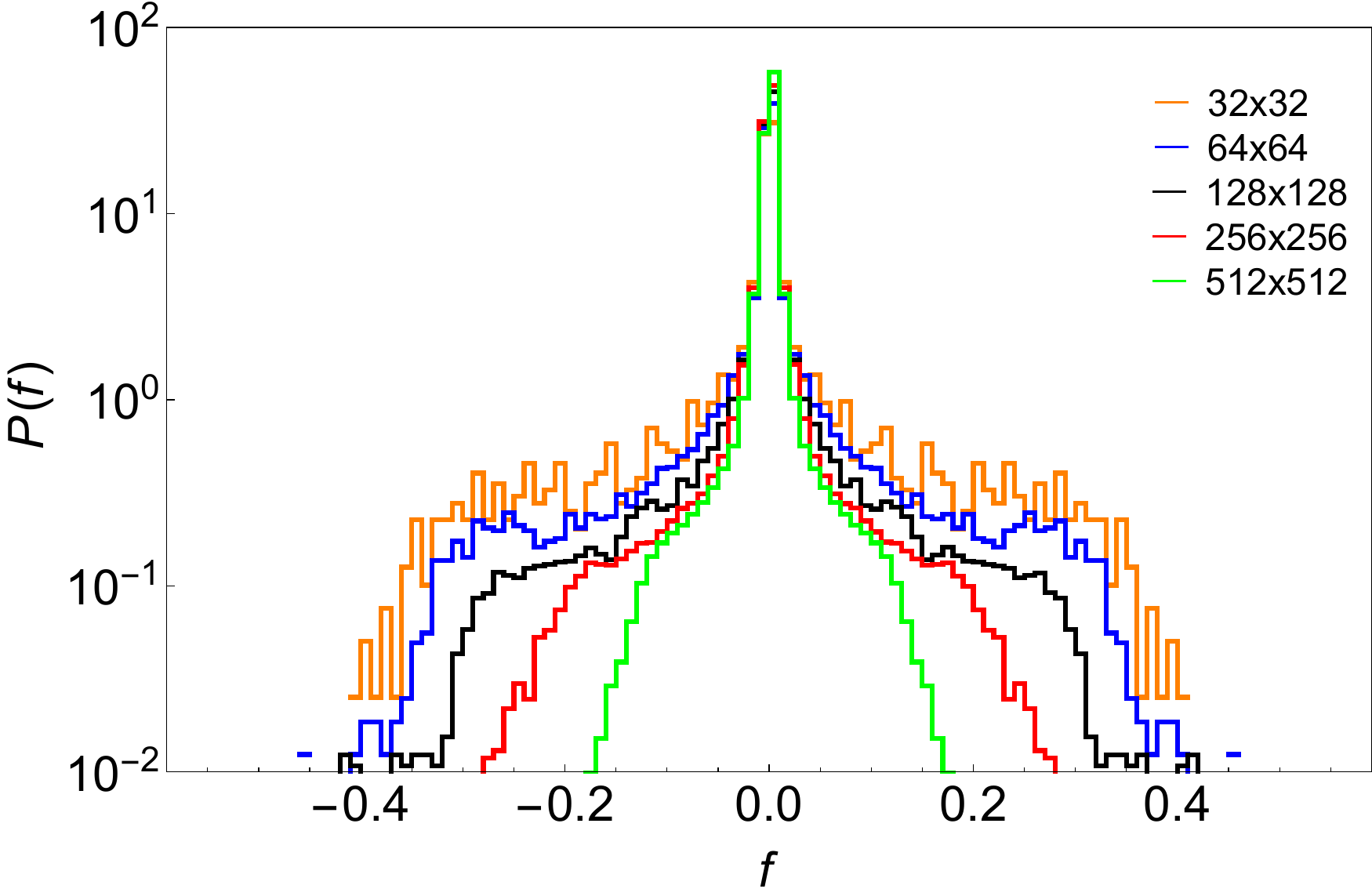}\\
    \includegraphics[width=4cm, height=4cm]{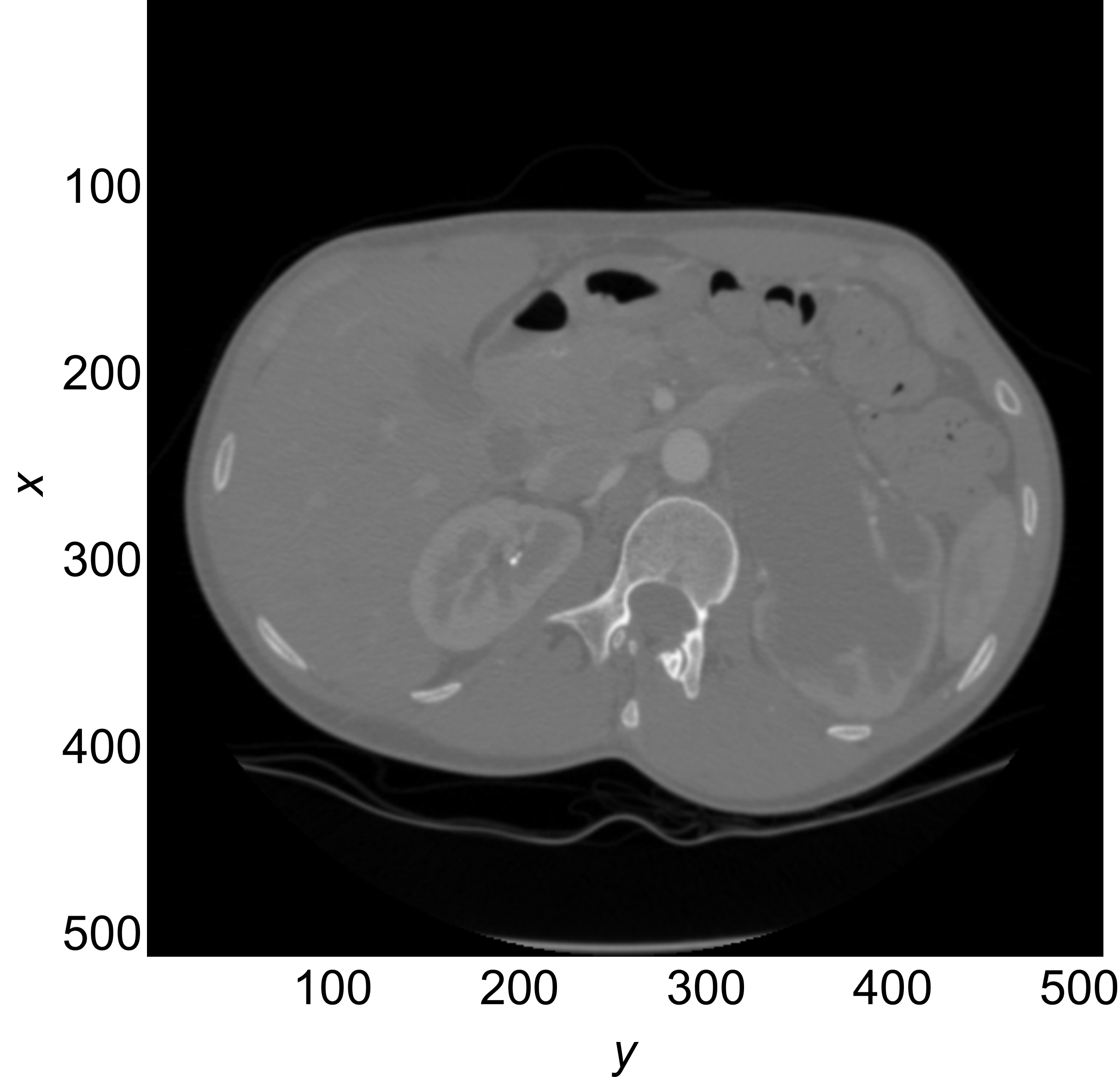}\includegraphics[width=4cm, height=4cm]{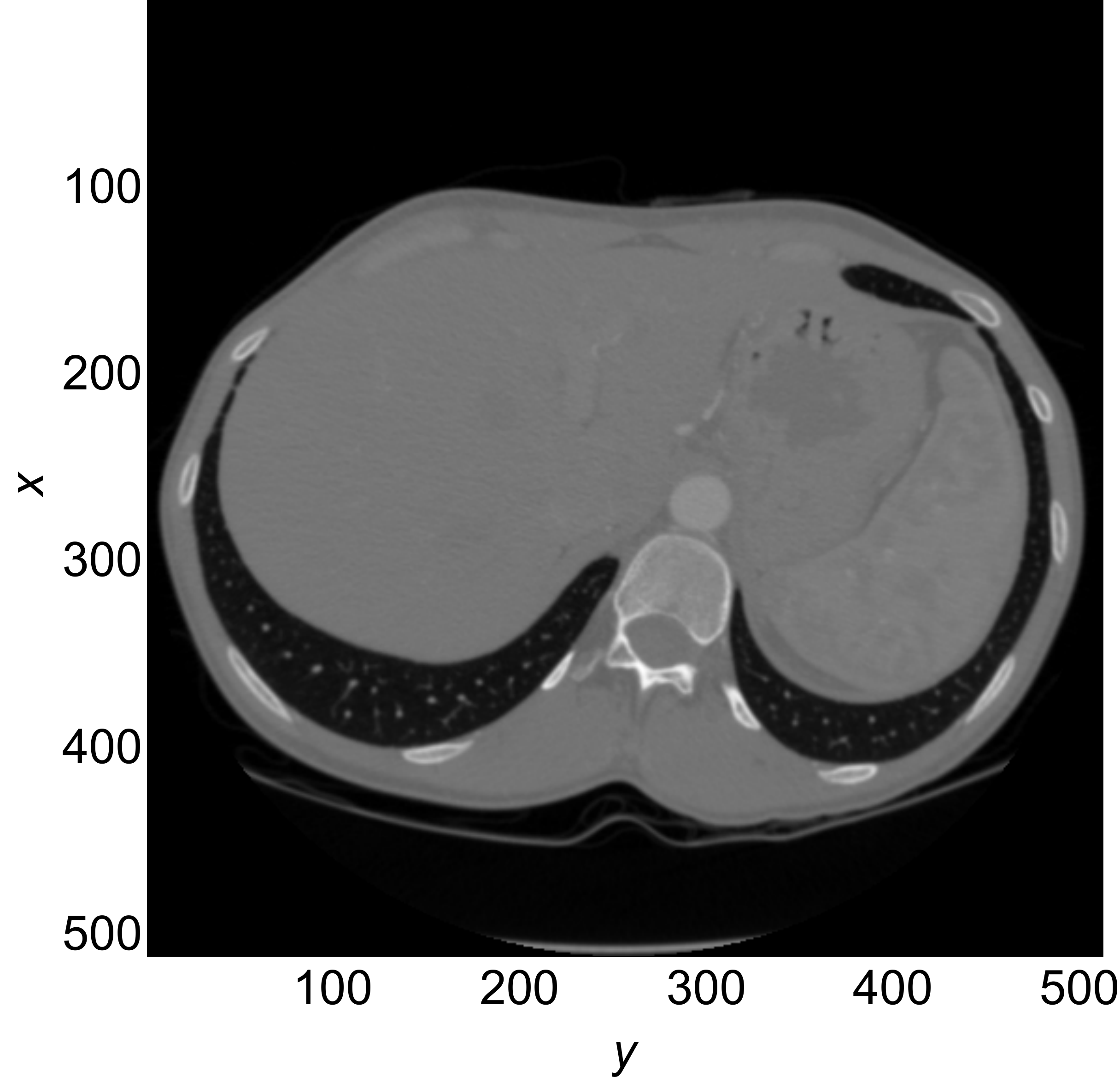}\\
    \vspace{0.7cm}
     \includegraphics[width=4cm, height=3.4cm]{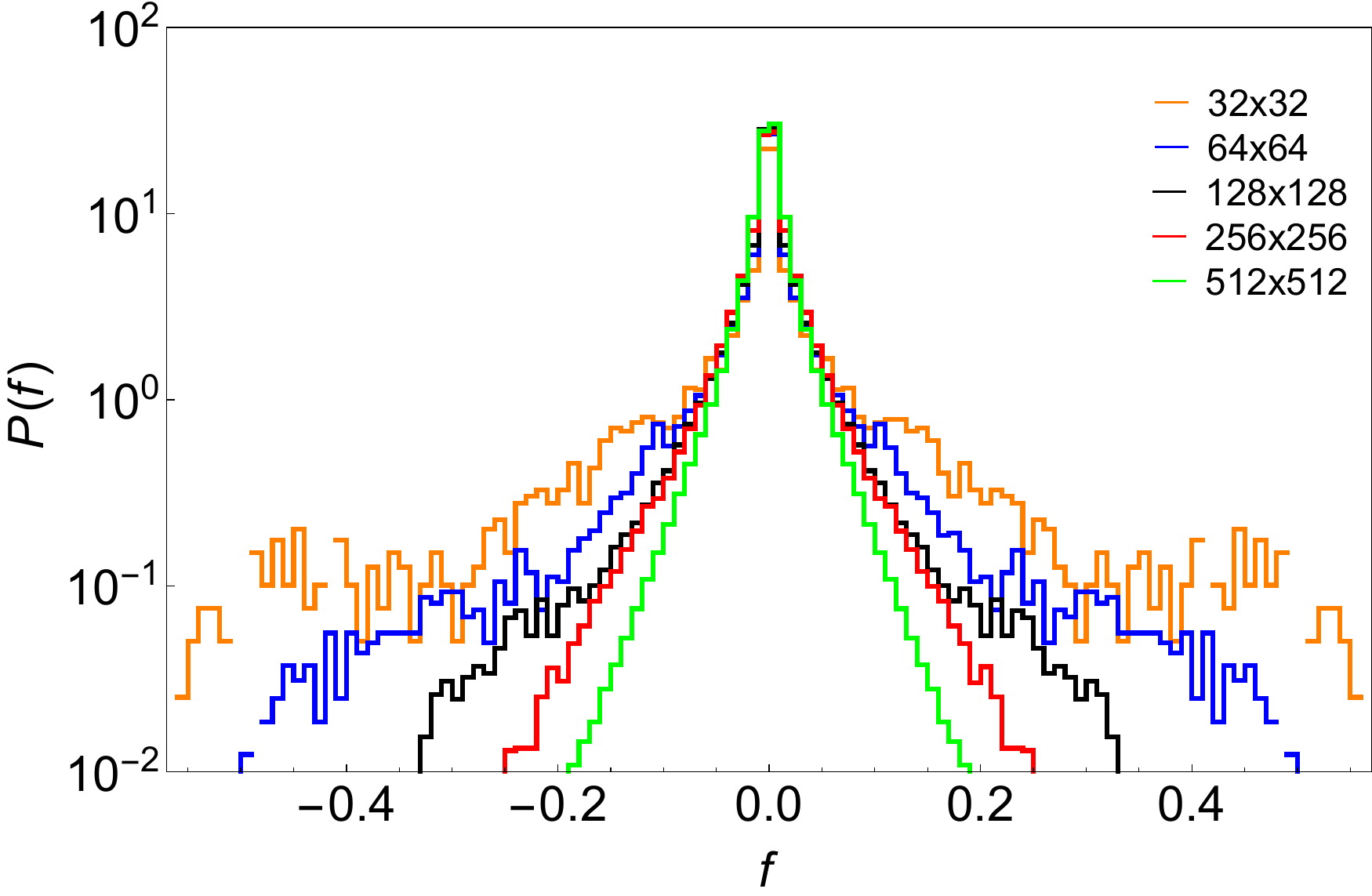}\includegraphics[width=4cm, height=3.4cm]{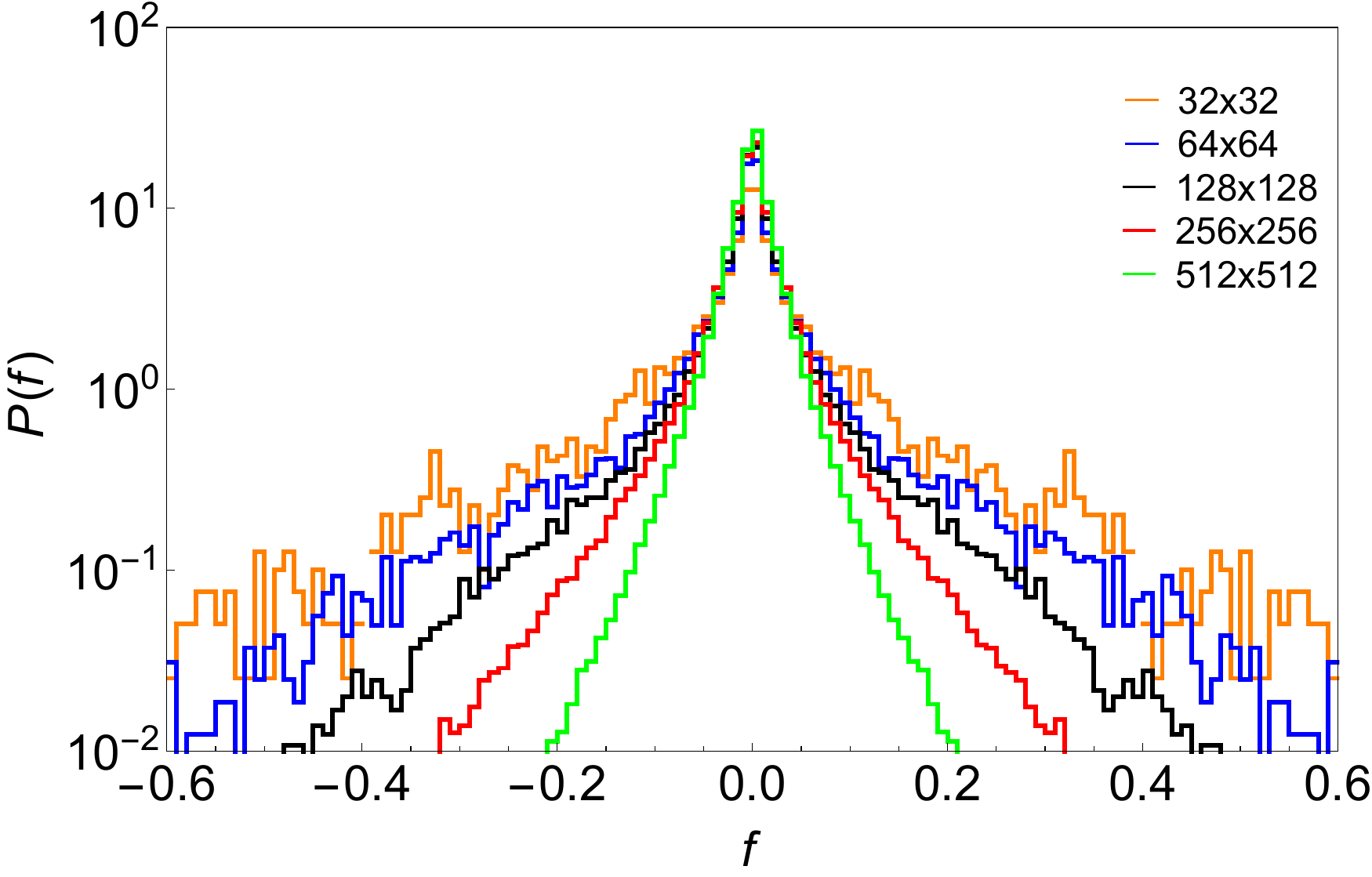}\\
    \includegraphics[width=4cm, height=4cm]{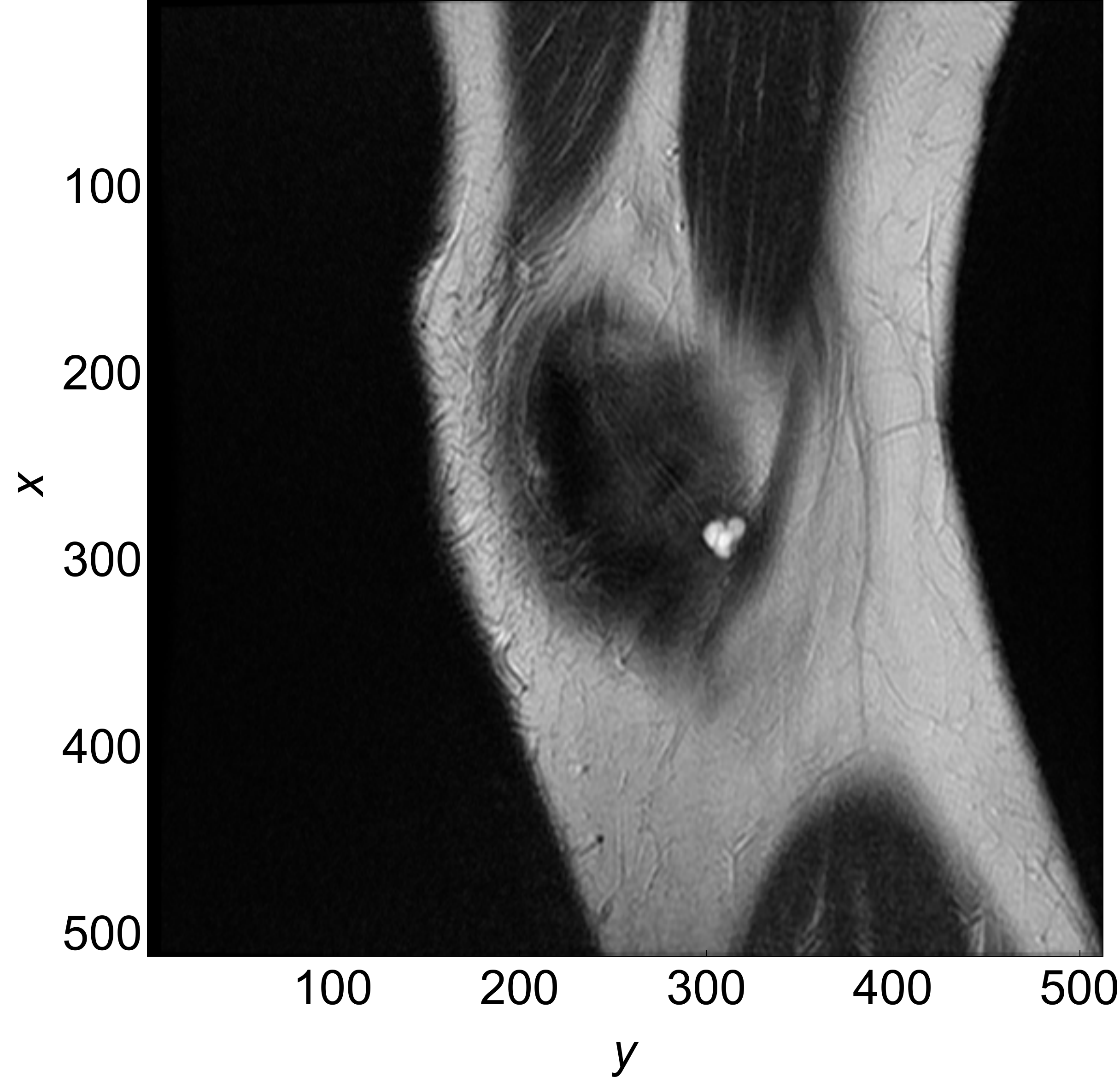}\includegraphics[width=4cm, height=4cm]{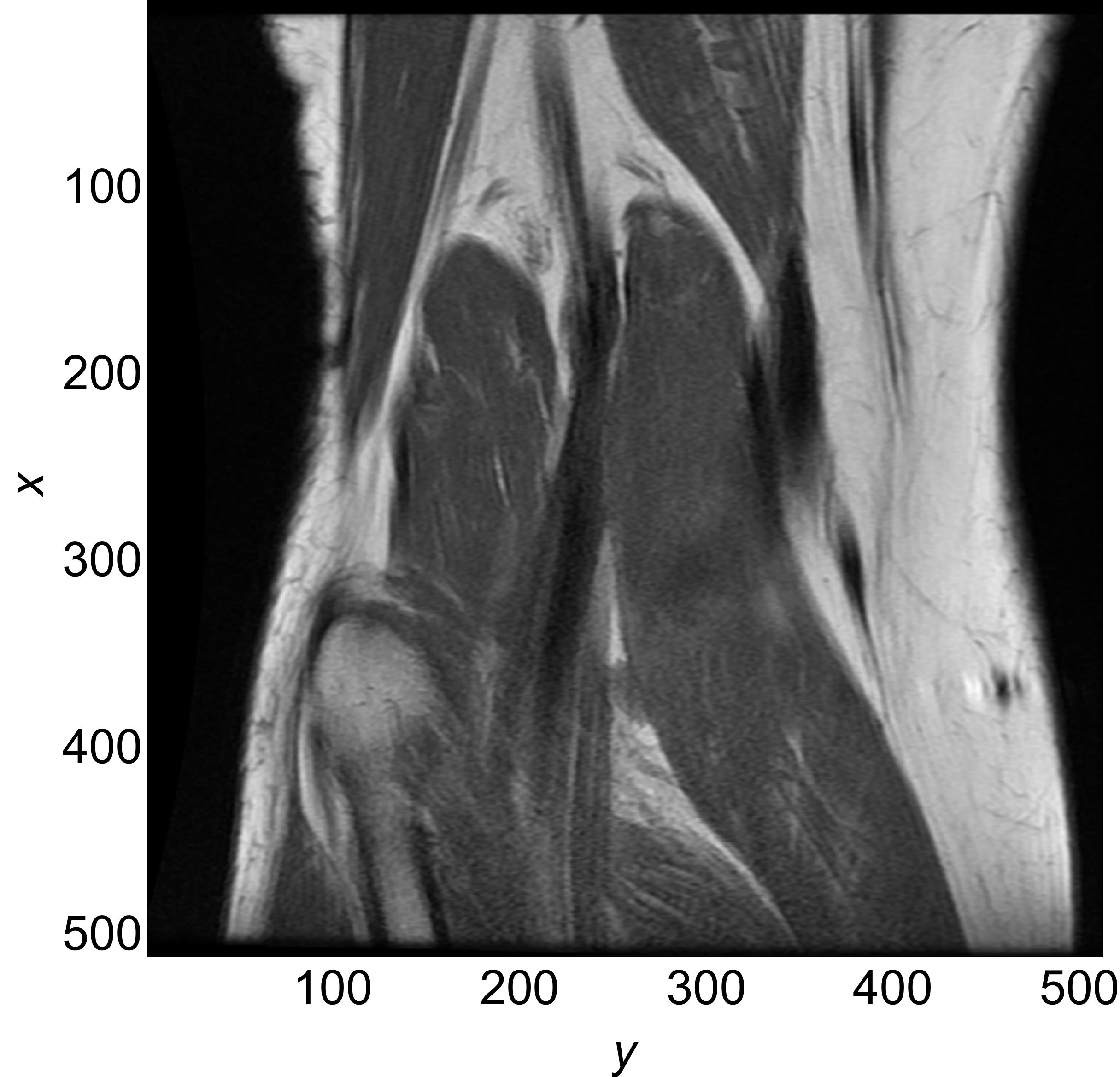}
  \end{center}
  \caption{Empirical distribution of the gradient measured on CT scans for  full resolution images ($512\times512$) and lower resolutions ($256\times256,128\times128,64\times64, 32\times32$). Upper panel: two different viewpoint of the abdomen; Lower panel: two different views of a knee.}
  \label{fig:organs}
\end{figure}

For the factor $P_0^{(\text{single})}(\bm{x})$, we will consider three different choices. The first one, which we will call \textit{interval} prior, corresponds to assuming a uniform measure on a generic support $[x^{(m)}_{i}, x^{(M)}_{i}]$ of the pixels, that is
\begin{equation}
P_{0,\rm{int}}^{(\text{single})}(\bm{x})=\prod_{i=1}^N\frac{\mathbb{I}_{x_i\in[x^{(m)}_{i},x^{(M)}_{i}]}}{x^{(M)}_{i}-x^{(m)}_{i}} \equiv \prod_{i=1}^N \Lambda_i(x_i) \,, 
\label{eq:interval_prior}
\end{equation}
where $\mathbb{I}_{A}$ denotes the indicator function of condition $A$. A second viable choice, which it is usually called the \emph{spike-and-slab} \cite{Ishwaran2005} or sparse prior,  is particularly useful in the reconstruction of images with extensive monochromatic background:
\begin{equation}
P_{0,\rm{sparse}}^{(\text{single})}(\bm{x})=\prod_{i=1}^N\left[s\delta(x_i)+(1-s)\Lambda_i(x_i)\right]\, .
\label{eq:sparse_prior}
\end{equation}
Here, the weighting factor  $s\in (0,1)$ is the sparseness parameter of the image and is  equal to the average fraction of background pixel within the image. Finally, in discrete binary tomography \cite{herman2012discrete,batenburg2011dart} one assumes that the two available colors are either black or white, corresponding to a region totally transparent or completely opaque, for which we will assign values $x_i = 0$ or $x_i = 1$, respectively. In this scenario, the single variable prior, that we will denote as the \emph{binary} prior,  takes the following simple form
\begin{equation}
P_{0,\rm{bin}}^{(\text{single})}(\bm{x})=\prod_{i=1}^N\left[s\delta(x_i)+(1-s)\delta(x_i-1) \right] \,.
\label{eq:bin_prior}
\end{equation}
The remaining factor $P_0^{(\text{pair})}(\bm{x})$ in the prior probability is supposed to favor images with certain features, such as a smooth change in the intensities of neighboring pixels. This accounts for the fact that real images possess local structure. A standard choice for $P_{0}^{(\text{pair})}$ is
\begin{equation}
  P_{0,\text{lap}}^{(\text{pair})}(\bm{x})\propto  e^{-\frac{J}{2}\bm{x}^T\cdot \bm{L}\cdot \bm{x}} \propto e^{-\frac J2  \sum_{i=1}^N\sum_{j\in\partial i}(x_i - x_j)^2}\,,
\label{eq:corr_priors}
\end{equation} 
where $\bm{L}$ is the Laplacian matrix of the nearest-pixels adjacency graph, $J$ a weight parameter, and $\partial i$ stands for the set of neighbors of pixel $i$. We will denote the prior in Eq. \eqref{eq:corr_priors} as $\ell_2$ smoothness, as it favors small norms of the finite differences gradient. Notice that this prior assumes a Gaussian profile for the probability density of the difference variables, which makes, in turn, the analytically treatment more amenable. Empirically, it turns out, at least for tomographic images, that the histogram of these auxiliary variables is far from being Gaussian distributed. Indeed,   Fig. \ref{fig:organs} shows the empirical frequency count of the gradient of the image, that is  $ P(f) \propto \sum_{i=1}^N \sum_{j\in\partial i}\delta(f-x_i + x_j)$, for a series of real CT scans.
 Interestingly, the empirical profiles do not depend much on the type of organ analyzed, but depend in a highly non trivial way on the coarse-graining of the image.   To capture these more realistic cases, we introduce the following {\em spike-and-slab} prior for neighboring pixel differences, which corresponds to finite differences partial derivatives: 
\begin{equation}
P_{0,\text{diff}}^{(\text{pair})}(x_i,x_j) \propto \rho\delta(x_i - x_j)+(1-\rho) e^{-\frac{\lambda}{2}(x_i-x_j)^2}\,,
\label{eq:difference_prior}
\end{equation} 
with $\rho\in[0,1]$ and $\lambda\geq 0$.  For brevity, we  will refer to this prior as $\ell_0$ \textit{smoothness}. As we will show below  this prior produces very accurate image reconstructions.

Once we have selected the prior distribution, the image is reconstructed by using the posterior distribution $P(\bm{x}|\bm{p})$. The main goal is to find an efficient way to extract the information of this posterior. EP was introduced to approximate posterior distributions, along with their marginals \cite{Opper2001,Minka2001,Opper2004}, for a large class of intractable probabilistic models. In the image reconstruction problem, the posterior distribution of images, given the projections, takes the functional form of a multivariate Gaussian with positive definite covariance matrix $\bm{\Sigma}$ and mean $\bm{\mu}$, times a product of univariate distributions $\psi_i(x_i)$:
\beeq{
  P(\bm{x}|\bm{p})=\frac1Z e^{-\frac12 (\bm{x}-\bm{\mu})^T \bm{\Sigma}^{-1} (\bm{x}-\bm{\mu})} \prod_i \psi_i(x_i)\,.
  \label{eq:posterior}
}
The multivariate Gaussian term takes into account the measurements, i.e. the likelihood term, times other interacting terms in the prior, if any. For instance, in the case of the prior given by Eq.~\eqref{eq:corr_priors}, the expressions for $\bm{\Sigma}$ and $\bm{\mu}$ are
\begin{equation}
\bm{\Sigma}^{-1} = \beta \bm{A}^T\bm{A} + J \bm{L}\, ,
\qquad
\bm{\mu} = \beta \bm{\Sigma} \bm{A}\bm{p}\, ,
\label{eq:post parameters}
\end{equation}
respectively. The set of  functions $\psi_i(x)$ account for non-Gaussian factors, such as density bounds between 0 and 1, $\ell_0$ sparsity, binary constraints, or non log-convex priors (explicit forms for other choices of priors can be found in Appendices \ref{app:A} and \ref{app:C}). To introduce the method of EP, we proceed as follows. Suppose that, to trade-off accuracy for solvability, we approximate Eq.~\eqref{eq:posterior} by replacing each $\psi_i(x_i)$ term by a normal density
\beeq{
\phi_i(x_i)=\frac{1}{\sqrt{2\pi b_i}}e^{-\frac{(x_i-a_i)^2}{2b_i}}\, . 
}
Then the new   posterior, denoted here as $Q_{\{\phi\}}(\bm{x})$, has the following  expression
\beeq{
  Q_{\{\phi\}}(\bm{x})=\frac1{Z_{Q_{\{\phi\}}}} e^{-\frac12 (\bm{x}-\bm{\mu})^T \bm{\Sigma}^{-1} (\bm{x}-\bm{\mu})}\prod_{i}\phi_i(x_i)
}
Note that $Q_{\{\phi\}}(\bm{x})$ is a multivariate Gaussian distribution for which it is easy to obtain the single variable marginals $Q_{\{\phi\}}(x_i)$, whence the value of each pixel can be inferred as the mean of its corresponding marginal. 

We are now left with the problem of choosing the mean and variance vectors, $\bm{a}=(a_1,\dots,a_N)$, $\bm{b}=(b_1,\dots,b_N)$ in order to best approximate the true posterior probability $P(\bm{x}|\bm{p})$ by using $Q_{\{\phi\}}(\bm{x})$. A seemingly reasonable form for the approximating factors $\phi_i$ would be the closest univariate Gaussians (in KL distance) to $\psi_i$. This approach, however, produces poor results in reconstruction. The EP algorithm improves strikingly on the MAP estimation by approximating not the $\psi_i$ measures themselves, but their effect on the full distribution. More precisely, we introduce the so-called tilted distribution for pixel $i$-th, $Q^{(i)}_{\{\phi\}}(\bm{x})$:
\beeq{
  Q^{(i)}_{\{\phi\}}(\bm{x})&=\frac1{Z_{Q^{(i)}_{\{\phi\}}}} e^{-\frac12 (\bm{x}-\bm{\mu})^T \bm{\Sigma}^{-1} (\bm{x}-\bm{\mu})}\psi_i(x_i)\prod_{j\neq i}\phi_j(x_j) \, .
}
One then chooses the parameters $(a_i,b_i)$ of the Gaussian distribution $\phi_i(x_i)$ such that the Kullback-Leibler distance $D_{\text{KL}}$ between $Q_{\{\phi\}}$ and $Q^{(i)}_{\{\phi\}}$ is minimized, that is
\beeq{
(a_i^\star,b_i^\star)=\text{arg min}_{(a_i,b_i)} D_{\text{KL}}[Q^{(i)}_{\{\phi\}}||Q_{\{\phi\}}] \, .
\label{eq:mmc}
}
It is straightforward to show \cite{Minka2001,Heskes2005,Opper2004} that the $D_{\text{KL}}$ minimization is equivalent to the following moment-matching condition
\beeq{
  \bracket{x_i}_{Q_{\{\phi\}}}=\bracket{x_i}_{{Q^{(i)}_{\{\phi\}}}}\,,\quad\quad \bracket{x^2_i}_{Q_{\{\phi\}}}=\bracket{x^2_i}_{Q^{(i)}_{\{\phi\}}}\,,
  \label{eq:apv}
}
where we have denoted $\bracket{\cdots}_\rho=\int dx \rho(x)(\cdots)$ for some distribution $\rho(x)$. In this way, Eq.~\eqref{eq:mmc} can be used as an iterative procedure until convergence is reached for every pair of parameters $(a_i,b_i)$. At convergence the value of each reconstructed pixel is determined by the formula
\beeq{
x_{i}^{*} = \bracket{x_i}_{{Q^{(i)}_{\{\phi\}}}}\,.
}

\section{Parameters estimation} 
\label{sec:parameters}
Unlike other methods used in image reconstruction, those based in Bayesian inference allow to estimate  fairly naturally the set of parameters of the model to obtain an optimal reconstruction. In our particular case, the set of parameters to infer depends on the particular choice of the prior distribution. To fix ideas let us consider, for instance, a prior distribution consisting on the binary prior, together with the $\ell_2$ smoothness prior, given by Eqs. \eqref{eq:bin_prior} and \eqref{eq:corr_priors}, respectively.  This choice then contains three parameters: the inverse variance noise distribution $\beta$, the weight of the Laplacian matrix $J$, and the sparseness parameter $s$. Using a free energy minimization procedure (further details can be found in Appendix \ref{app:D}), all of them can be inferred within the EP iteration process. Actually, the sparseness parameter $s$, can be easily estimated using a gradient-descent scheme, while the other two can be approximated via the Expectation Maximization technique \cite{dempster1977maximum}. Indeed, if $\bm{x}_{\rm{EP}}^{(t)}$ represents the EP reconstruction at iteration step $t$,  then one obtains the following iterative set of  of equations to render  the best estimates of  these three parameters
\begin{equation}
\begin{aligned}
s^{(t+1)} & = s^{(t)} + \eta \frac{\partial F^{(t)}_{\text{EP}}}{\partial s}\, ,\\
\beta^{(t+1)} & = \frac{M}{( \bm{A}\bm{x}_{\rm{EP}}^{(t)}-\bm{p})^T( \bm{A}\bm{x}_{\rm{EP}}^{(t)}-\bm{p})}\, ,\\
J^{(t+1)} & = \frac{N}{\bm{x}_{\rm{EP}}^{(t)} \bm{L} \bm{x}_{\rm{EP}}^{(t)}}\, .
\end{aligned}
\end{equation}
Here $F_\text{EP}^{(t)}$ is the EP free energy (see Eq.\eqref{eq:F_EP} in Appendix \ref{app:D}) at iteration step $t$ evaluated using $\bm{x}^{(t)}_{\text{EP}}$, while $\eta$ is a relaxation parameter of the gradient descent algorithm. For other choices of prior distribution, the estimation of the corresponding parameters can be carried out in a  similar manner.

\section{Results}
\label{sec:results}

\subsection{Results for phantom images}
To estimate the goodness of the prior distribution choice, we compare our performances against three reconstruction methods commonly used in the literature: TV (for $\ell_1$ smoothness), Quadratic Programming (QP) (for $\ell_2$ smoothness) and, for binary reconstruction only, the BP algorithm. Experiments consist in the reconstruction of \emph{ensembles} of phantom images in different noise and measurements regimes. The noise distribution is considered to be known for QP, TV and BP (since there is no clear strategy to estimate it within the algorithms), while, for the implementations of EP, we estimate $\beta$ as described in Sec. \ref{sec:parameters}.

\begin{figure}
	\begin{center}
		\includegraphics[width=\columnwidth]{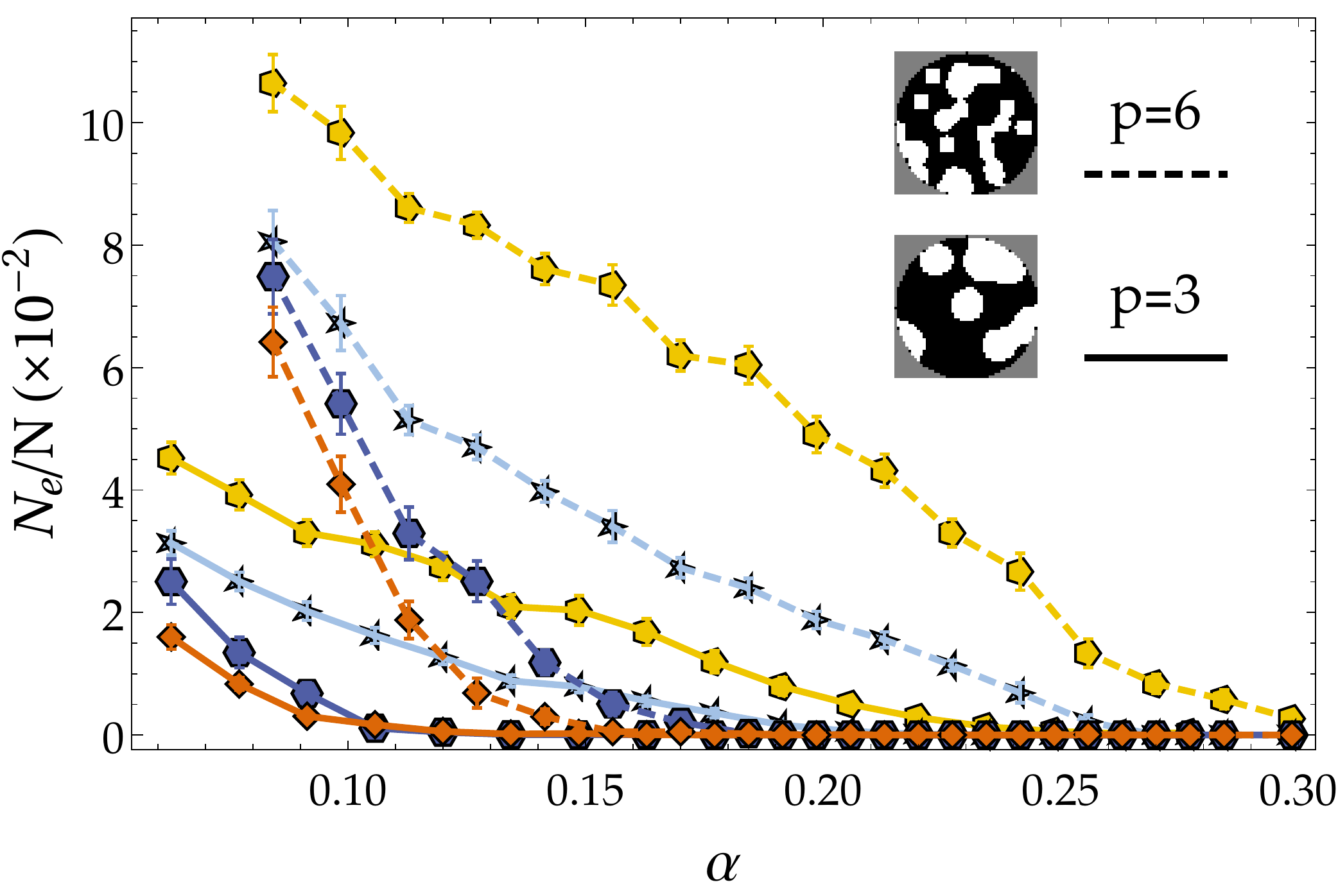}\\
		\includegraphics[width=\columnwidth]{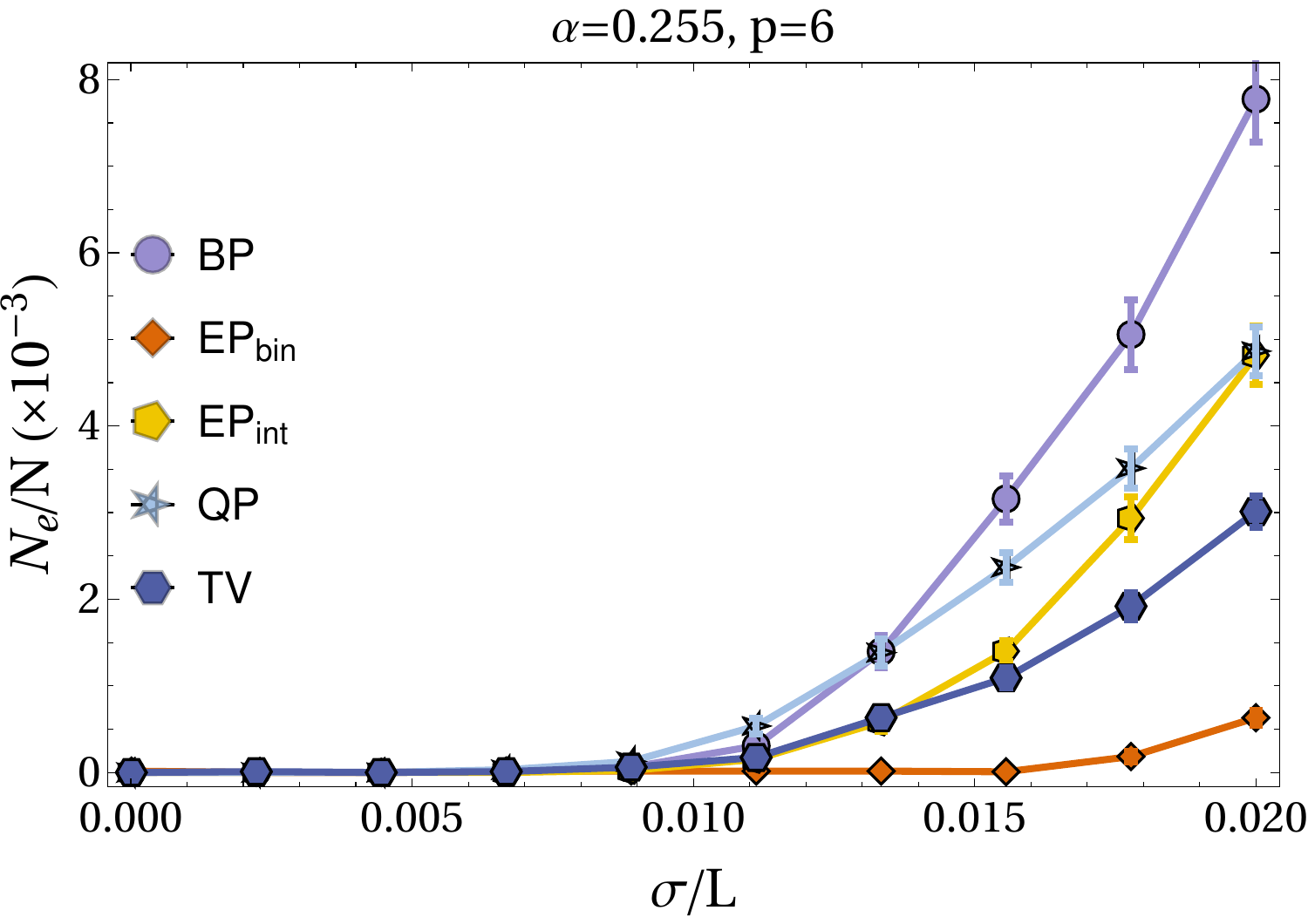}
	\end{center}
	\caption{Fraction of error pixels versus sampling rate $\alpha$ (upper panel) and noise-to-signal ratio $\sigma/L$ (lower panel) with $\sigma$ the standard deviation for the Gaussian noise for binary images. Each point corresponds to the results after averaging over 50 randomly generated phantoms, while the error bars are one standard deviation from the average.}
	\label{fig1}
\end{figure}

Synthetic phantoms represent light patches, or clusters, in a circular black background (as in the inset of Fig.~\ref{fig1}) that are generated as follows: starting from a black colored image of dimension $L \times L$, we color uniformly at random $p^2$ pixels of ``white'' that we will be used as centroids of a Gaussian filter of width $\sim 1/p$. For binary tomography only, we binarize the resulting images (further details in the generation of phantom images are provided in App. \ref{app:G}). By tuning $p$ we can control the structure and complexity of the phantoms since it has been found empirically that the number of pixels in the boundary between the light regions and the background, and thus the dimension of the patches, scales roughly linearly with $p$ \cite{Gouillart2013}. For each chosen value of $p$, we generate a set of images differing in the choice of the seeds. After a measuring step, in which we mimic a realistic acquisition process of projections, we try to reconstruct the phantoms. To simulate the noisy regime we add a random variable with Gaussian distribution $\mathcal{N}(0,\sigma)$ to each component of the measurement vector $\bm{p}$, for $\sigma = \beta^{-1/2}$. Finally, to quantify the accuracy of the reconstructed image $\bm{x}^\star$ compared to the real one $\bm{x}$, we introduce two metrics: for binary images, we use  the number of wrongly assigned pixel $N_e$, while for non-binary images, we estimate the average $\ell_2$ norm of the difference between the original image and the reconstructed one, that is, $E_2 = ||\bm{x}-\bm{x}^*||_2/N$, where $N$ is the number of the pixels within the circular regions. All images used for these experiments have sizes $L = 50$ \footnote{Note that the number of pixels to reconstruct, $N$, is not $L^2$, since only the values inside the circular region are to be inferred, thus leaving $N=1959$.}, while the values of $p$ will be specified case by case. 

The results for binary reconstruction obtained with EP, TV and QP are shown in Fig.~\ref{fig1}.  The subscript \texttt{bin} (resp. \texttt{int}) refers to the use of the binary (resp. interval) prior in the posterior distribution. The top panel depicts the dependence of the fraction of errors $N_{e}/N$ upon the sampling rate $\alpha \equiv M/N$. The inset  show a typical realization of such phantoms for synthetic random phantoms generated using $p = 3$ and $p = 6$. The reconstruction error is non-negligible up to a certain value of $\alpha$ above which perfect reconstruction is reached. Recall that the smaller the value of $\alpha$, the less number of measurements $M$ we need to achieve a good performance. 
The lower panel, on the other hand, hosts the results for a fixed value of $\alpha=0.255$ and $p=6$, and shows the fraction of errors as a function of the noise-to-signal ratio, $\sigma/L$. Notice that for this value of $\alpha$ the reconstruction error is zero for all the methods in the noiseless case (for $\sigma$ very small) and then increases for non-negligible value of the noise. As we can see, the EP algorithm always achieves a lower error fraction when using a binary prior as it outperforms any other reconstruction method in the noiseless and noisy scenarios. \\

\begin{figure}
	\begin{center}
		\includegraphics[width=\columnwidth]{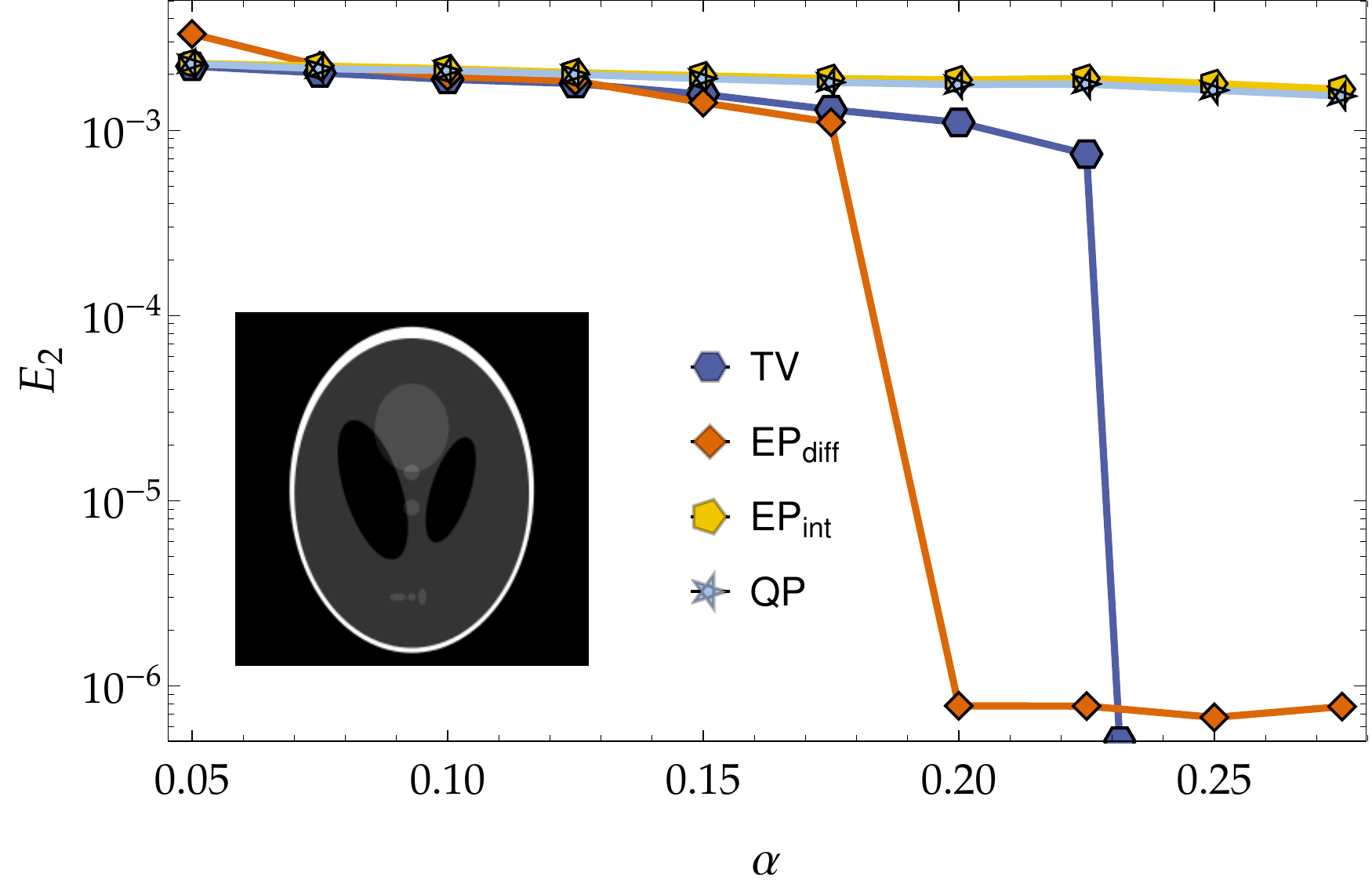}\\
		\includegraphics[width=\columnwidth]{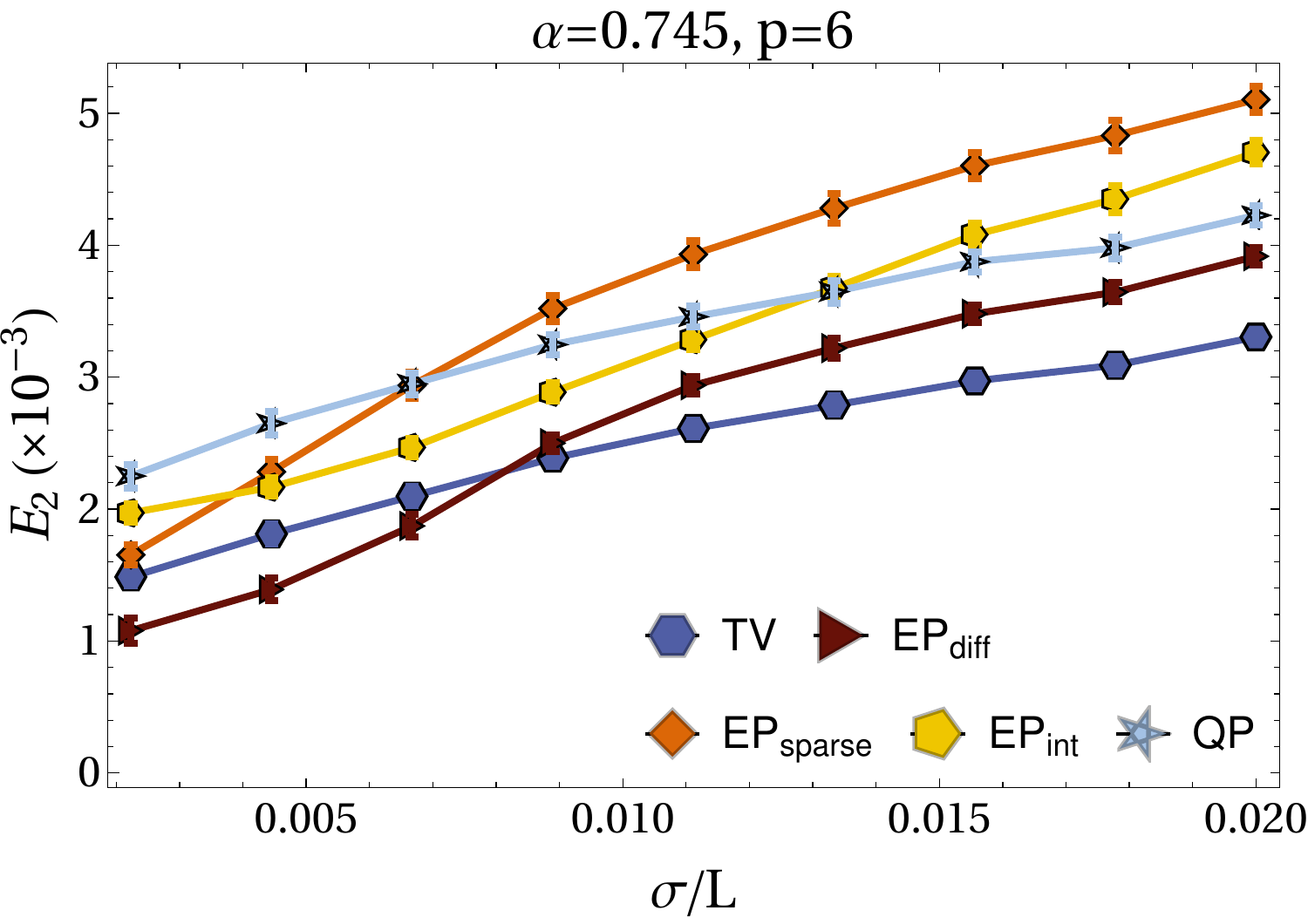}
	\end{center}
	\caption{$E_2$ error for continuous images as a function of the sampling rate $\alpha$ for the Shepp-Logan phantom (upper panel), and as a function of noise-to-signal ratio $\sigma/L$ using synthetic random phantoms, for fixed $\alpha$ and $p$ (lower panel).}
		
	\label{fig2}
	
\end{figure}

For non-binary images, we first tested our method using the Shepp-Logan phantom \cite{shepp_fourier_1974}, a well known benchmark synthetic image representing a 2D section of a simplified human head. Experiments are performed in a noise-free setup and for $L = 80$.  In the implementation of the EP algorithm applied to this image we have tried out  three different priors, namely the  interval, the  sparse and the difference priors, given by   Eqs.~\eqref{eq:interval_prior}, \eqref{eq:sparse_prior}, and \eqref{eq:difference_prior}, respectively. The upper panel of Fig.~\ref{fig2} shows the reconstruction error of all the algorithms under study as a function of the sampling rate $\alpha$. Empirically we found that, for non-binary images, a perfect reconstruction is reached whenever $E_2 \simeq 10^{-4}$, since for such reconstructed images there are not discernible corrections. According to our findings, EP with the difference prior reaches this threshold error, and thus achieves a perfect reconstruction, for $\alpha$ in  the interval $(0.18, 0.20)$, considerably before the other algorithms. To further benchmark the versatility of EP algorithm, we have studied a noisy case in which we apply EP, QP and TV to an ensemble of synthetic non-binary images for $p = 6$ . The lower panel of Fig. \ref{fig2} depicts the behavior of $E_2$, averaged over $50$ synthetic images of $N = 1959$ pixels, as a function of the noise-to-signal ratio.  As we can see from this figure, EP with a difference variables prior outperforms the other algorithms for moderate values of noise (low values of $\sigma$) and then perform very similarly to TV (that computes the best reconstructed images) for larger value of the noise. A similar behavior is also found as a function of $\sigma/L$ for smaller values of $\alpha$. It is important to keep in mind, however, that when using TV and QP the noise distribution is assumed to be known, while in the EP approach it is an additional parameter to be inferred. This means that EP with the difference prior performs closely to TV even when less information about the measurement setup is available.

\subsection{Results for real tomographic images}
We report here the results of QP, TV and the two implementations of EP algorithm for non-binary pixels (with \textit{interval} and $\ell_0$ smoothness  priors) on the reconstruction of four real computed tomography (CT) scans: a mouse's head and three images of a human head differing in the acquisition plane. The index \textit{$i$} associated with each $\textit{CT\_head\_i}$ image refers the position of the scanner with respect to the neck of the patient (the smaller the index \textit{$i$} the closer to the neck). For these experiments, the original high resolution images are rescaled to a smaller size of $100\times100$ pixels and the measuring process has been simulated by our acquisition algorithm in the noiseless regime.

\begin{figure}
	\begin{center}
		\includegraphics[width=4.1cm,height=4.1cm]{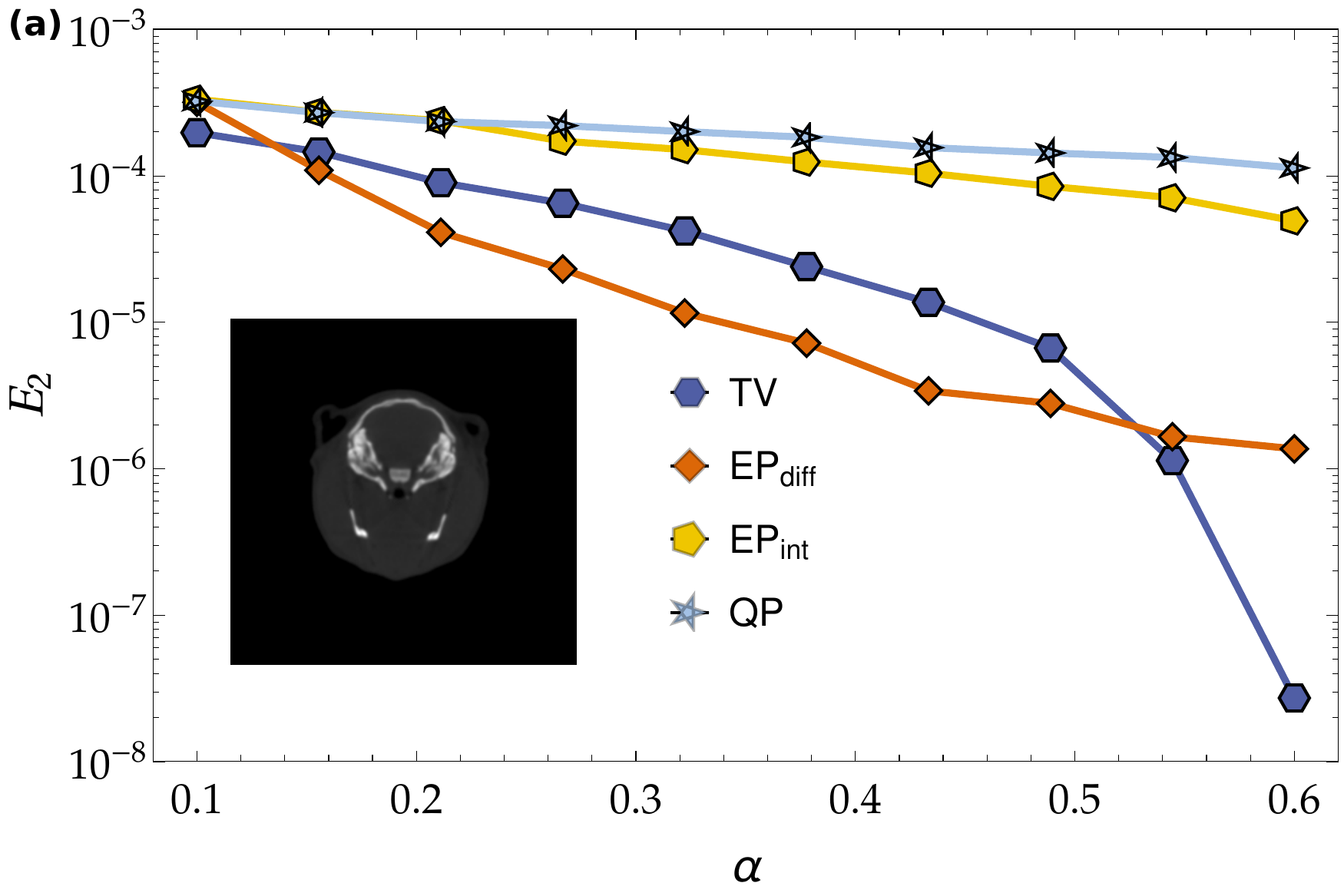}
		\includegraphics[width=4.1cm,height=4.1cm]{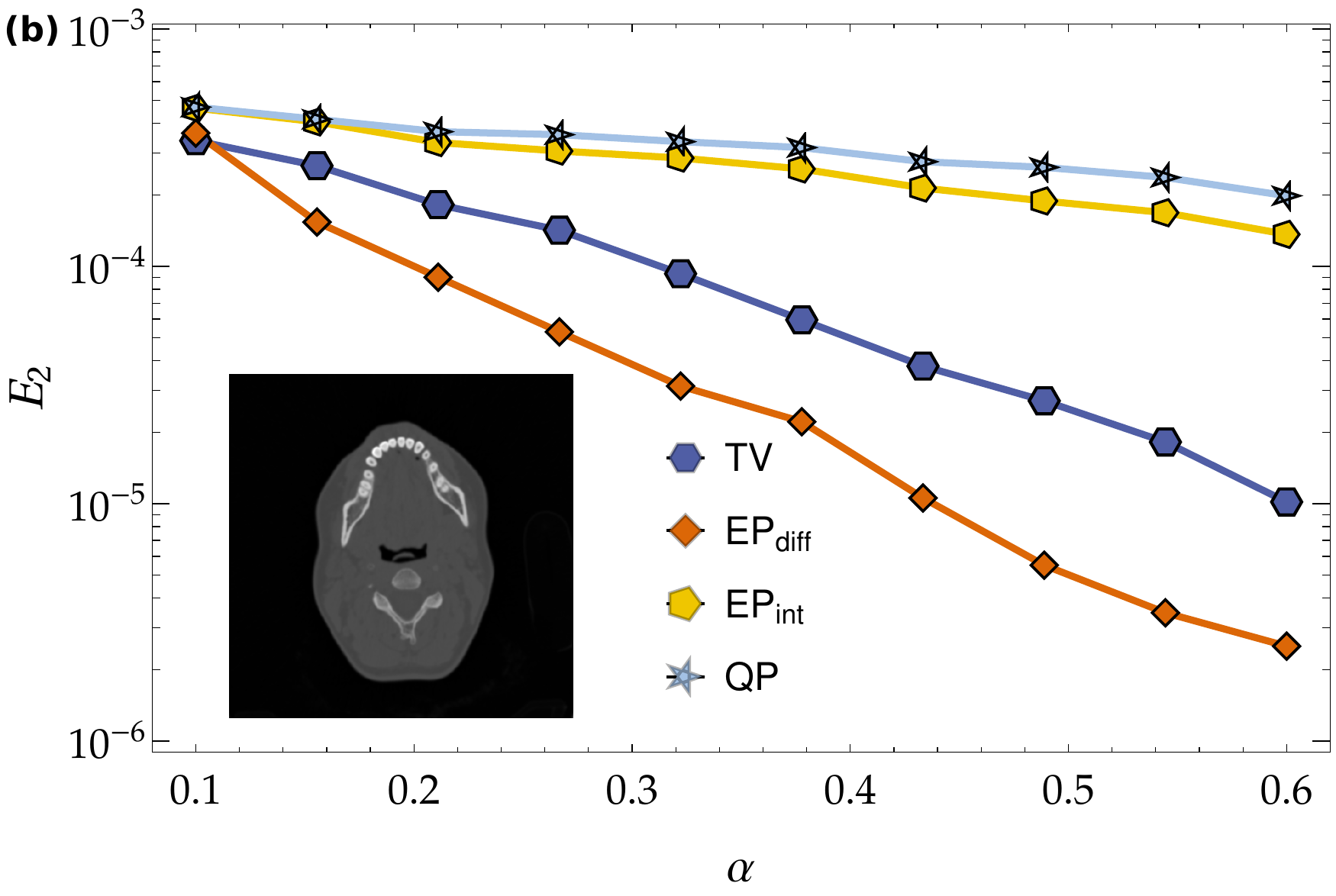}\\
		\includegraphics[width=4.1cm,height=4.1cm]{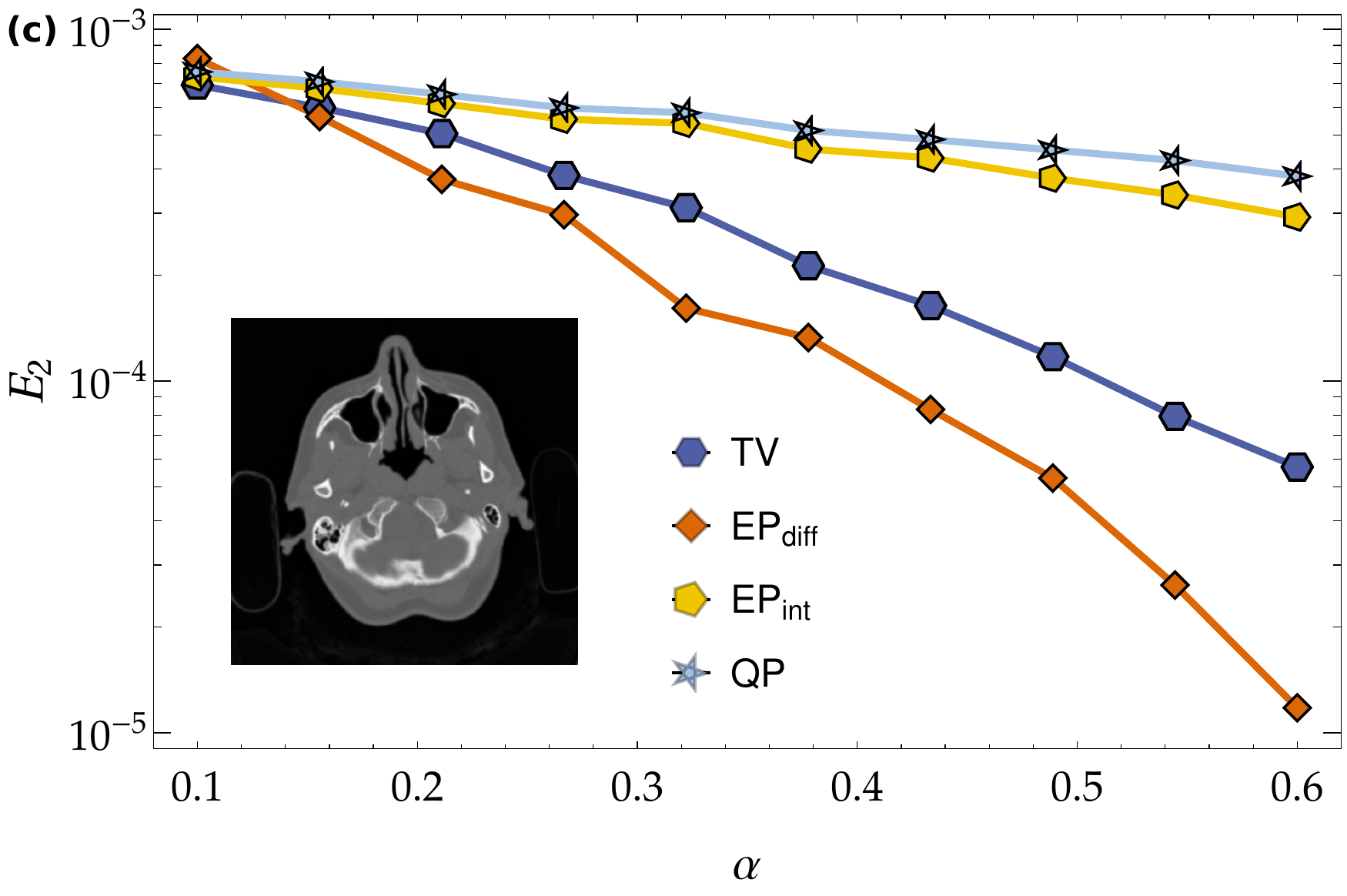}
		\includegraphics[width=4.1cm,height=4.1cm]{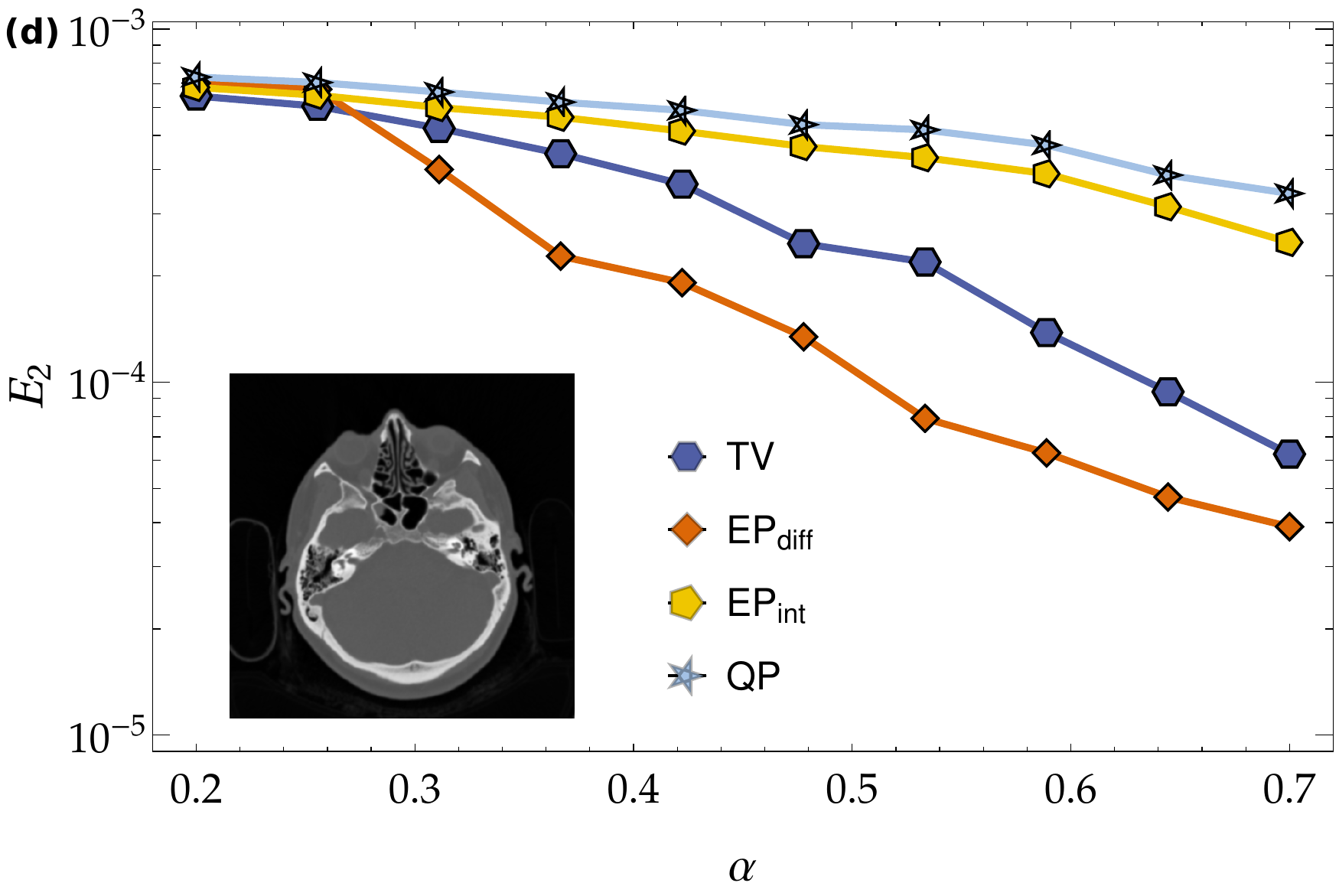}
	\end{center}
	\caption{$E_2$ error for real tomographic images as a function of the sampling rate $\alpha$}
	\label{fig_realtomo}
\end{figure}

\begin{figure*}
	\begin{center}
		\includegraphics[height=2\columnwidth, width=10cm, angle = 270]{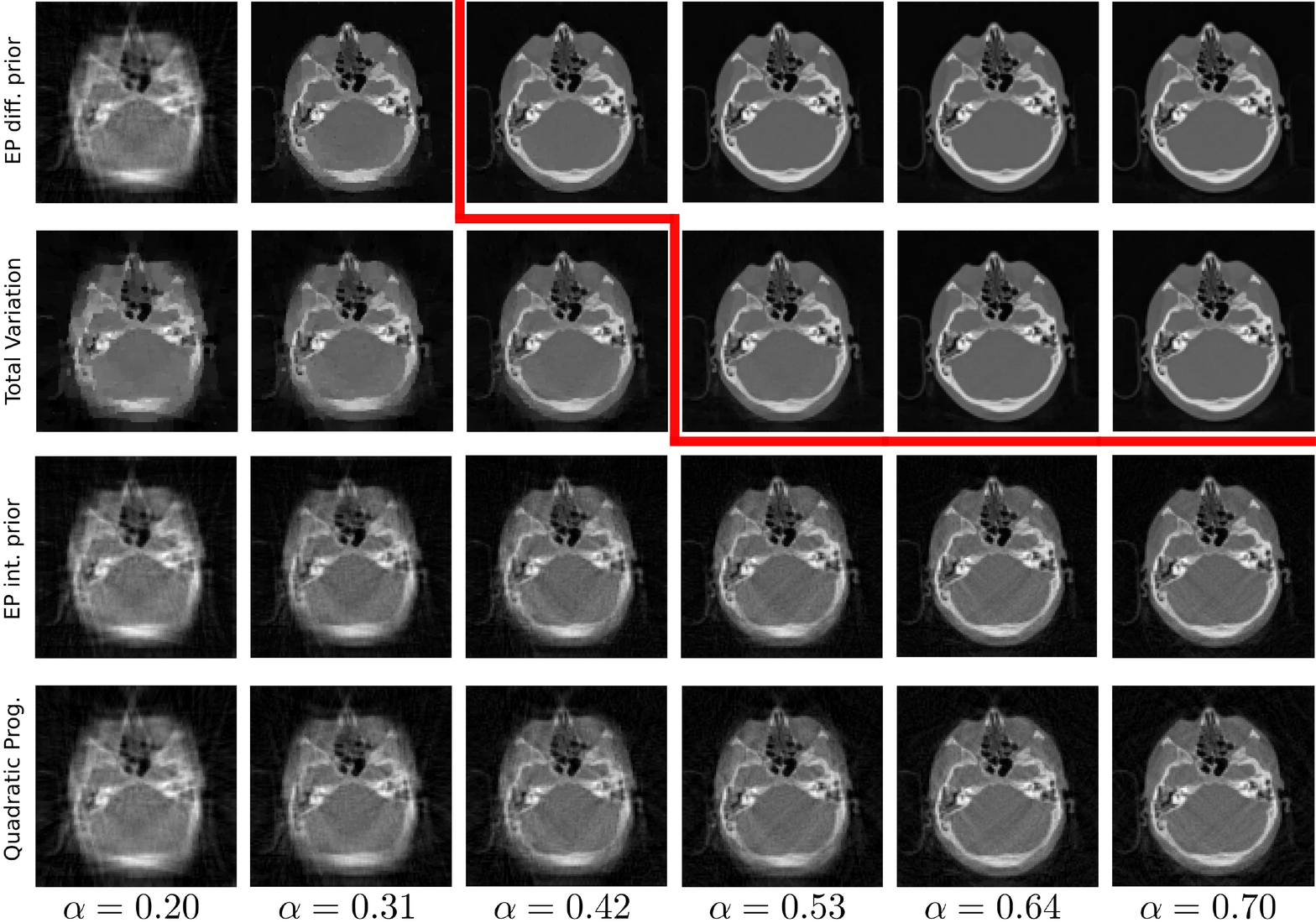}
	\end{center}
	\caption{Table containing the reconstructions of $\textit{CT\_head\_100}$ using EP$_\texttt{diff}$, TV, EP$_ \texttt{int}$ and QP as a function of the sampling rate $\alpha$. Top right region, delimited by the red line, contains the perfect image reconstructions.}
	\label{fig_cthead100}
\end{figure*}

In Fig.~\ref{fig_realtomo} we plot the reconstruction error \textit{$E_{2}$} as a function of $\alpha \in \left[0.1, 0.6\right]$ for the following cases: (\textbf{a}) the mouse's skull , (\textbf{b}) $\textit{CT\_head\_38}$ ,  (\textbf{c}) $\textit{CT\_head\_80}$, and  (\textbf{d}) $\textit{CT\_head\_100}$.  Rather remarkable, contrary to what we observed in the Shepp Logan phantom,  here the $E_{2}$ error of all algorithms decreases rather smoothly as we include more measurements. However, the transition to a perfect reconstruction regime is reached only by TV and EP with difference prior. This clearly indicates that   including pairwise interactions in the prior is certainly advantageous when dealing with tomographic images. It is worth noting that perfect reconstruction is reached by EP with the  $\ell_0$ smoothness  prior at $\alpha = 0.42$, while the TV algorithm needs more measurements to achieve  the same result, namely $\alpha=0.53$. Hence, one again, EP surpasses the other algorithms.  This is illustrated in Fig. \ref{fig_cthead100} where we show several reconstruction of  $\textit{CT\_head\_100}$ as a function of $\alpha$. The red line marks the boundary between perfectly reconstructed images (right region) and less accurate reconstructions (left region). TV and EP with  the $\ell_0$ smoothness prior reach the perfect reconstruction at $\alpha = 0.53$ and $\alpha = 0.42$, respectively, while EP with interval prior and QP need more measurements (that is, larger values of $\alpha$) to achieve an error-less reconstruction.

\subsection{Phase-type diagram of perfect/imperfect reconstruction}
\begin{figure}

\centering
\includegraphics[width=\columnwidth]{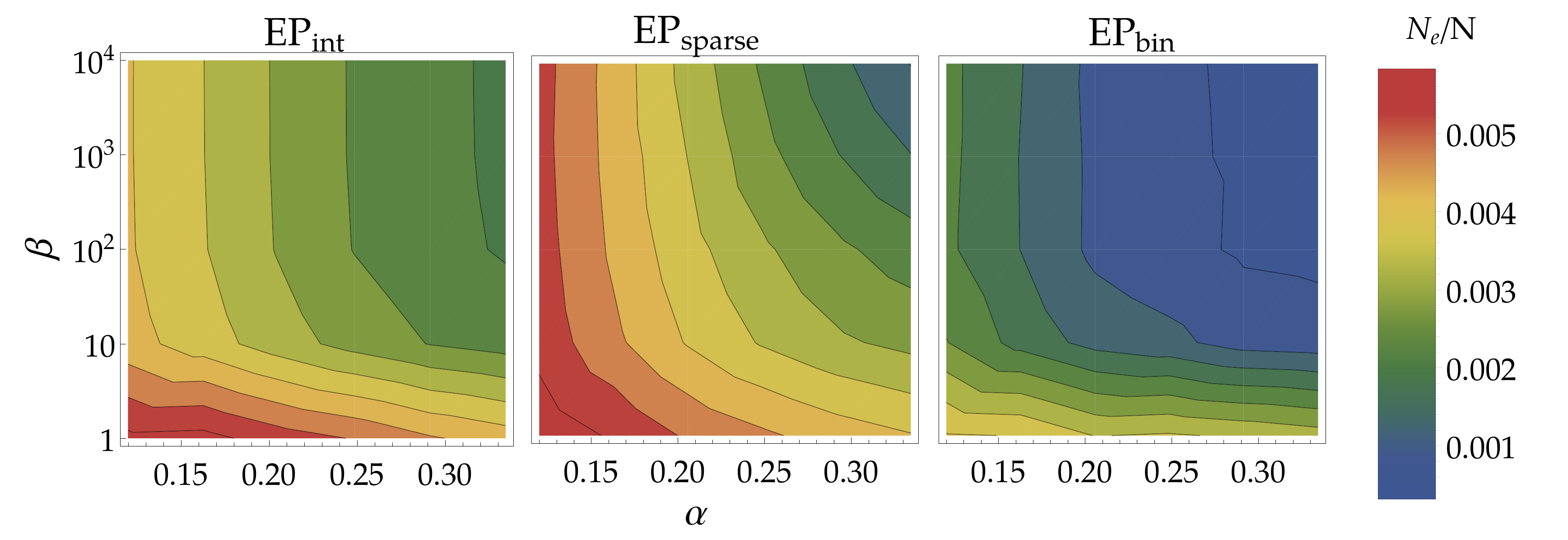} \\
\includegraphics[width=\columnwidth]{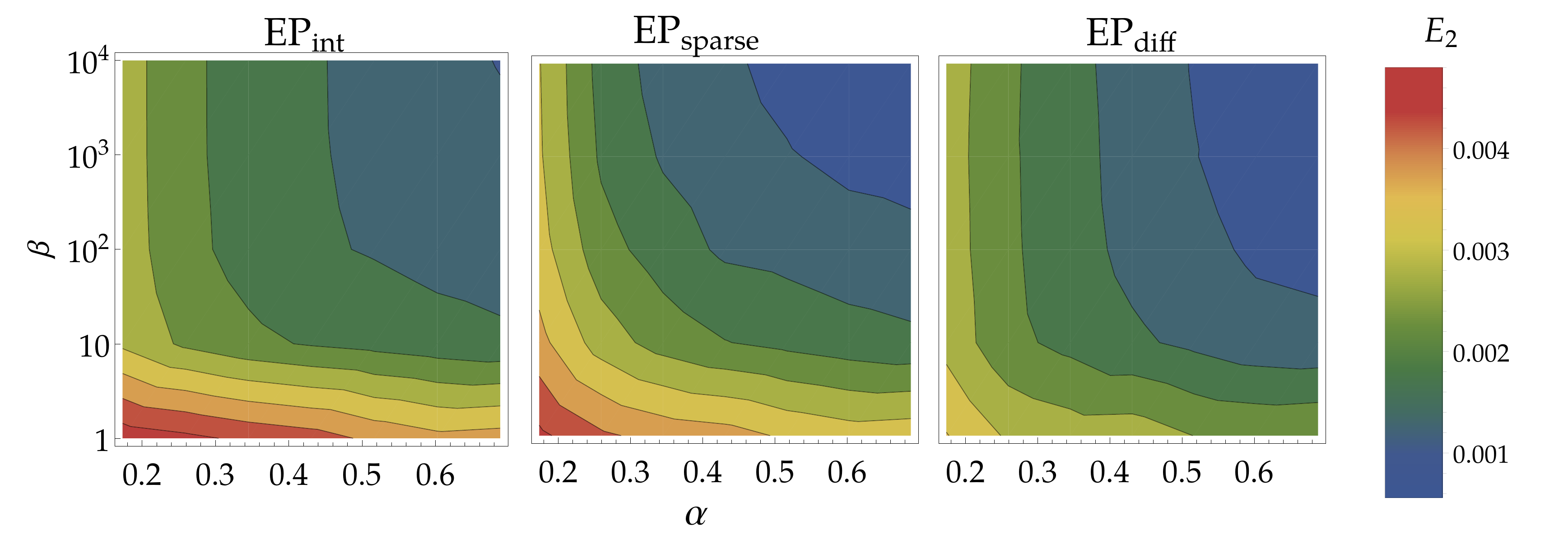}
\caption{Reconstruction error in the $(\alpha,\beta)$ plane, comparing the performance of the various priors we have tested, for binary (upper panel) and continuous (lower panel) images. As explained in the text,  EP\textsubscript{int}, EP\textsubscript{bin}, EP\textsubscript{sparse}, and EP\textsubscript{diff} correspond to interval prior, binary prior, sparseness prior and difference-variable prior, respectively.} 
\label{fig3}
\end{figure}

To characterize the performances of all implementations of EP we show here the  perfect/imperfect reconstruction diagrams in the $(\alpha,\beta)$-plane. The tested images are the ones used for the lower panels of Figures \ref{fig1} and \ref{fig2}. Results are shown in Figure \ref{fig3} for both, binary (upper panel) and non-binary (lower panel) reconstructions. As we can see from the plots, the implementations of EP that reach the best performances (lower reconstruction errors) for both binary and gray-scale images, are the ones with the binary prior and the difference prior respectively. Intuitively, this confirms that inference performance is strongly tied to the closeness of the prior distribution to the correct statistics of the target ensemble of images. 

\section{Discussion}
\label{sec:discussion}
We have shown how to address the problem of reconstructing tomographic images by including non-standard prior information about the image, normally resulting in non-log concave prior weight functions. The reconstruction itself can be performed by the EP algorithm. EP is able to encode, within a Bayesian framework, empirical information about the statistics of the treated variables, using \emph{ad-hoc} prior distributions that are rather difficult or even impossible to cope with standard tools. The results presented here employ prior knowledge both about single pixel and the differences of nearest-neighbors pixels intensities. For sake of simplicity, the prior distribution over these auxiliary variables does not take into account the spatial localization of the pixels but, in principle, EP can treat even this more specific case. For instance, one can exploit a collection of a certain class of tomographic images as a training set for the statistics of each difference variable, with the resulting histograms then encoded as prior distributions for future inferences. Notice that the treatment of difference variables is possible within the EP framework because it involves a linear transformation of the pixel intensities. From a more general perspective, one can think of extending the same formalism to any linear transformation of the pixel variables. 

We have compared the performances of EP to the ones of standard convex optimization techniques and, only in the case of binary tomography, to the BP algorithm. Binary tomography results show that, using a binary prior, a perfect reconstruction is possible  for synthetic images  even within the limited-data regime, outperforming any other algorithms adopted here. In the case of non-binary images, EP performs remarkably well when using the difference prior, carrying a clear improvement in the inference when compared to EP implementations with interval and sparse prior. This suggests that when dealing with a more specific type of reconstruction problem, a drastic improvement can be attained by employing a prior that describes well the specific subclass of target images. With respect to other techniques, EP reconstructions present more accurate reconstruction with respect to TV ones in the case of the Shepp-Logan phantom, real tomographic images and for synthetic images affected by noise. 

It is worth pointing out that EP running time is dominated by a matrix inversion per iteration that requires $\mathcal{O}\left(N^{3}\right)$ operations when using only priors over pixels. Unfortunately, this limits the size of the images to be reconstructed and thus prevent the direct applicability of these methods to real tomography where a high resolution images are generally required. For larger images, a hierarchical iterative approach can be implemented, in which (cheaper) reconstruction at smaller resolutions is used as initial state for more expensive reconstruction at larger resolution.

Not only EP is able to approximate well the posterior distributions of the models presented here but it also provides a powerful tool to estimate the parameters of the model and to have access to the properties of the noise distribution affecting the data that are, in real case scenarios, unknown. We remark that, in contrast to EP, neither TV nor QP are able to infer the parameters of the optimization and thus several runs are needed before reaching the best reconstruction. Moreover, the model presented here deals with additive noise but the multiplicative noise regime can be faced by EP using a slightly different formulation of the likelihood.

\begin{acknowledgments}
IPC is grateful to the project DGAPA-UNAM-PAPIIT IA103417 and thanks HuGeF and Polito for their hospitality. AP, AB, and APM acknowledge funding from INFERNET, a European Union's Horizon 2020 research and innovation program under the Marie Sk\l odowska-Curie grant agreement No 734439. RDHR is grateful to CONACYT for financial support. ICP and RDHR thank Arnulfo Martinez for interesting discussions and clarifications about image reconstruction on medical images.  The medical images used in the present work were generated during the research carried out under the grant UNAM-DGAPA-PAPIIT IN108615 and IN110616, as well as CONACYT Problemas Nacionales 2015-01-612.
\end{acknowledgments}

\appendix

\section{Details of the EP algorithm for pixels probability distributions} 
\label{app:A}
We derive here the equations of the EP algorithm for the interval prior.  Other cases, except for the difference variables prior that will be treated separately later on, are completely analogous. For the former, let us first write  the approximate posteriors $Q$ and $Q^{(i)}$ as follows:
\beeq{
Q(\bm{x}|\bm{p})&=\frac{1}{Z_Q}e^{-\frac{1}{2}(\bm{x}-\bm{\mu}^{(i)})^T\cdot\bm{\Sigma}^{-1}_{(i)}\cdot(\bm{x}-\bm{\mu}^{(i)})}\frac{e^{-\frac{(x_i-a_i)^2}{2b_i}}}{\sqrt{2\pi b_i}}\,,\\
Q^{(i)}(\bm{x}|\bm{p})&=\frac{1}{Z_Q^{(i)}}e^{-\frac{1}{2}(\bm{x}-\bm{\mu}^{(i)})^T\cdot\bm{\Sigma}^{-1}_{(i)}\cdot(\bm{x}-\bm{\mu}^{(i)})}\frac{\mathbb{I}_{x_i\in [x_i^{(m)},x_i^{(M)}]}}{x_i^{(M)}-x_i^{(m)}}\,,
}
where we have defined the following matrices and vectors:
\beeq{
\bm{\Sigma}^{-1}_{(i)}&=\beta \bm{A}^T\bm{A}+J\bm{L}+\bm{B}^{(i)}\,,\\
\bm{\mu}^{(i)}&=\bm{\Sigma}_{(i)}\cdot(\beta \bm{A}^T\cdot\bm{p}+\bm{B}^{(i)}\cdot\bm{a})\,,\\
\bm{B}^{(i)}&=\text{diag}(b_1^{-1},\ldots, b_{i-1}^{-1}, 0, b_{i+1}^{-1},\ldots,b_N^{-1})\,.
}

The minimization of the Kullback-Leibler distance between $Q$ and $Q^{(i)}$ yields the moment matching conditions of Eq.~\eqref{eq:mmc}, which when solved for the parameters $\{(a_i,b_i)\}_{i=1}^N$ result in:
\beeq{
b_i&=\left(\frac{1}{\bracket{x^2_i}_{Q^{(i)}}-\bracket{x_i}^2_{Q^{(i)}}}-\frac{1}{\Sigma_{ii}}\right)^{-1}\,,\\
a_i&=b_i\left[\bracket{x_i}_{Q^{(i)}}\left(\frac{1}{b_i}+\frac{1}{\Sigma_{ii}}\right)-\frac{\mu_i}{\Sigma_{ii}}\right]\, .
\label{eq:parameters a,b}
}
To shorten notation we have defined $\Sigma_{ii}\equiv (\bm{\Sigma}_{(i)})_{ii}$, whereas $\bracket{x_i}_{Q^{(i)}}$ and $\sigma^2_{Q^{(i)}}\equiv\bracket{x^2_i}_{Q^{(i)}}-\bracket{x_i}^2_{Q^{(i)}}$ have the following expressions

\begin{widetext}
\beeq{
\langle x_{i}\rangle_{Q^{\left(i\right)}} &=\mu_{i}+\frac{\mathcal{{N}}\left(\frac{x_{i}^{(m)}-\mu_{i}}{\sqrt{\Sigma_{ii}}}\right)-\mathcal{{N}}\left(\frac{x_{i}^{(M)}-\mu_{i}}{\sqrt{\Sigma_{ii}}}\right)}{\Phi\left(\frac{x_{i}^{(M)}-\mu_{i}}{\sqrt{\Sigma_{ii}}}\right)-\Phi\left(\frac{x_{i}^{(m)}-\mu_{i}}{\sqrt{\Sigma_{ii}}}\right)}\Sigma_{ii}\, ,\\
\sigma^2_{Q^{(i)}}\, 
&=\Sigma_{ii}\left[1+\frac{\frac{x_{i}^{(m)}-\mu_{i}}{\Sigma_{ii}}\mathcal{{N}}\left(\frac{x_{i}^{(m)}-\mu_{i}}{\sqrt{\Sigma_{ii}}}\right)-\frac{x_{i}^{(M)}-\mu_{i}}{\sqrt{\Sigma_{ii}}}\mathcal{{N}}\left(\frac{x_{i}^{(M)}-\mu_{i}}{\sqrt{\Sigma_{ii}}}\right)}{\Phi\left(\frac{x_{i}^{(M)}-\mu_{i}}{\sqrt{\Sigma_{ii}}}\right)-\Phi\left(\frac{x_{i}^{(m)}-\mu_{i}}{\sqrt{\Sigma_{ii}}}\right)}-\left(\frac{\mathcal{{N}}\left(\frac{x_{i}^{(m)}-\mu_{i}}{\sqrt{\Sigma_{ii}}}\right)-\mathcal{{N}}\left(\frac{x_{i}^{(M)}-\mu_{i}}{\sqrt{\Sigma_{ii}}}\right)}{\Phi\left(\frac{x_{i}^{(M)}-\mu_{i}}{\sqrt{\Sigma_{ii}}}\right)-\Phi\left(\frac{x_{i}^{(m)}-\mu_{i}}{\sqrt{\Sigma_{ii}}}\right)}\right)^{2}\right] \, ,
}
\end{widetext}
with   definitions
\beeq{
\Phi(x)=\frac{1}{2}\left[1+\text{erf}\left(\frac{x}{\sqrt{2}}\right)\right]\,,\quad\quad \mathcal{N}(x)=\frac{1}{\sqrt{2\pi}} e^{-\frac{x^2}{2}} \, .
}
\section{Fast computation of the update equations}
\label{app:B}
The aforementioned moment matching conditions, which  appear in any implementation of the EP algorithm,  require inverting $\bm{\Sigma}_{(i)}^{{-1}}$, since $a_i$ and $b_i$ depend explicitly on the $i$-th diagonal element of $\bm{\Sigma}$. On top of that, a direct implementation of the procedure described so far would involve performing this inversion for \emph{each pixel}, thus resulting in an algorithm that scales as $\mathcal{O}(N^4)$ per iteration step. This is a consequence of the fact that inverting an $N\times N $ matrix scales as $N^3$. However, let us define
\begin{equation}
\begin{aligned}
\overline{\bm{\Sigma}}^{-1} & = \beta \bm{A}^T\bm{A} + J \bm{L} + \bm{B}\, ,\\
\overline{\bm{\mu}} & = \overline{\bm{\Sigma}} \left( \beta \bm{A}^T\bm{p} + \bm{B}\bm{a} \right)\,,
\end{aligned}
\end{equation}
with $\bm{B}$ a full diagonal matrix of elements $B_{ii}=b_i^{-1}$. Using these quantities, we can reduce the computational cost of the EP algorithm, since now  we are able to compute the necessary elements for applying the moment matching conditions with a \emph{single matrix inversion per iteration step}. After some basic algebra, we notice  that we can express $\Sigma_{ii}$ and $\mu_i$ as
\begin{equation} \label{eqs:mod parameters}
\begin{aligned}
\Sigma_{ii} & = \frac{\overline{\Sigma}_{ii}}{1-\overline{\Sigma}_{ii}/b_i}\, ,\\
\mu_i & = \frac{\overline{\mu}_i - \frac{a_i}{b_i}\overline{\Sigma}_{ii}}{1-\overline{\Sigma}_{ii}/b_i}\, .
\end{aligned} 
\end{equation}
Even though this still requires $\mathcal{O}(N^4)$ operations, we  have replaced $N$ matrix inversions per iteration step for $N$ arithmetic operations, as found in Eqs.~\eqref{eqs:mod parameters}. As a side effect, we should also consider that this improvement in performance limits us to a \emph{parallel} updating scheme for the values of $\bm{a}$ and $\bm{b}$, instead of a sequential one.

\section{Prior on difference variables}
\label{app:C}
Let us introduce a set of difference variables $f_{ij} = x_{i} - x_{j}$ for $j \in \partial i$, along with the pixels variables $\bm{x}$, having a prior distribution as in Eq. \eqref{eq:difference_prior} on the main text. The joint posterior probability of intensities and differences is written as
\beeq{
P(\bm{x},\bm{f}|\bm{p})&=\frac{1}{Z} e^{-\frac{\beta_1(\bm{A}\bm{x}-\bm{p})^T(\bm{A}\bm{x}-\bm{p})}{2}} e^{-\frac{\beta_2}{2}\sum_{i\sim j} (x_i-x_j -f_{ij})^2} \\
&\times\prod_{i}\mathbb{I}_{x_i\in [x_i^{(\text{m})},x_i^{(\text{M})}]}\prod_{i\sim j} \left[\rho\delta (f_{ij})+(1-\rho)e^{-\frac{\lambda}{2}f_{ij}^2}\right]\,,
}
where $i\sim j$ stands for summing over distinct pairs of neighboring pixels, and we expect to take the limit $\beta_{2} \rightarrow \infty$. Let  $E$ be the number of difference variables and let us introduce the vector $\bm{t}=\begin{pmatrix}\bm{x}\\\bm{f}\end{pmatrix}$. Further, let us define the following $(M+E)\times (N+E)$ matrix  $\bm{S}_{(M+E)\times (N+E)}$  written in block form:
\beeq{
\bm{S}_{(M+E)\times (N+E)}=\begin{pmatrix}
\sqrt{\beta_1}\bm{A}&\bm{0}_{M\times E}\\
\sqrt{\beta_1}\bm{R}_{E\times N}&-\sqrt{\beta_2}\bm{I}_{E\times E}\
\end{pmatrix}\,.
}
Here, $\bm{R}$ is a matrix whose entries  are given by $R_{i\sim j,i}=1$ and $R_{i\sim j,j}=-1$. Then the posterior can be rewritten as:
\begin{widetext}
\beeq{
P(\bm{t}|\bm{p})&\propto e^{-\frac{1}{2}(\bm{S}\bm{t}-\tilde{\bm{p}})^T(\bm{S}\bm{t}-\tilde{\bm{p}})}\prod_{i = 1}^{N}\mathbb{I}_{t_i\in [x_i^{(\text{m})},x_i^{(\text{M})}]}\prod_{i=N+1}^{N+E} \left[\rho\delta (t_{i})+(1-\rho)e^{-\frac{\lambda}{2}t_{i}^2}\right]\,,
}
\end{widetext}
with $\tilde{\bm{p}}=\begin{pmatrix}\sqrt{\beta_1}\bm{y}\\\bm{0}\end{pmatrix}^T$. According to the EP approximation scheme, we approximate each single-variable non Gaussian prior via $\mathcal{N}(a_{i}, b_{i})$ whose parameters are determined through the update equation in Eq. \eqref{eq:parameters a,b}. Notice that it depends on the form of the tilted distribution that, for this choice of priors, reads
\begin{widetext}
\beeq{
	Q^{(i)}(\bm{t}|\bm{p})&=\frac{1}{Z_Q^{(i)}}e^{-\frac{1}{2}(\bm{t}-\bm{\mu}^{(i)})^T\cdot\bm{\Sigma}^{-1}_{(i)}\cdot(\bm{t}-\bm{\mu}^{(i)})}\left\{\begin{array}{ll}
		\mathbb{I}_{t_i\in [x_i^{(\text{m})},x_i^{(\text{M})}]}&i\leq N\\
		\rho\delta (t_{i})+(1-\rho)e^{-\frac{\lambda}{2}t_{i}^2}&i>N
	\end{array}\right.\,,
}
with 
\beeq{
	\bm{\Sigma}_{(i)}^{-1}&=\bm{S}^T\bm{S}+\bm{B}^{(i)}\,,\quad 
	\bm{\mu}^{(i)}=\bm{\Sigma}_{(i)}\left( \bm{S}^T\tilde{\bm{p}}+\bm{B}^{(i)}\bm{a}\right)\,.
}
\end{widetext}
\section{Estimate of the parameters}
\label{app:D}
Unlike quadratic programming or total variation, the method presented here allows to estimate  the optimal values of the parameters used in the posterior distribution. For instance, in the case of binary images, one of the most important parameters is the sparseness  $s$, used in the binary prior (as defined in Eq. \eqref{eq:bin_prior} in the main text). To find its optimal value, denoted here $s^\star$, we first introduce thes EP free energy \cite{Opper2004}
\begin{equation} \label{eq:F_EP}
F_{\text{EP}} = (N - 1) \log Z_{Q} - \sum_{i=1}^N \log Z_{Q^{(i)}}\, ,
\end{equation}
where $Z_Q$ is the partition function of the approximating distribution, $Q(\bm{x}|\bm{p})$, and $Z_{Q^{(i)}}$  the corresponding one for the tilted distribution $Q^{(i)}(\bm{x}|\bm{p})$. To find the optimal value of $s$ we use the gradient descent method,
\begin{equation}
s^{(t+1)} = s^{(t)} - \eta \frac{\partial F_{\text{EP}}}{\partial s}\, ,
\end{equation}
with $\eta>0$ a relaxation parameter. Upon convergence, this scheme yields an estimation of $s^\star$ which  corresponds to a local minimum, thus  assuming that the real image sparseness minimizes the EP free energy. Given that our binary reconstructions did show a low fraction of errors under many circumstances once $s^\star$ was inferred, this seems to be indeed the case. The value of $\rho$ and $\lambda$ in the difference variables prior, appearing in Eq. \eqref{eq:difference_prior}, can be inferred with this same technique.

Due to the difficulties of computing the partial derivatives of $F_{\text{EP}}$ with respect of $\beta$ and $J$, the optimal values of these parameters are instead approximated using the Expectation Maximization (EM) algorithm \cite{dempster1977maximum}, which works as follows. Given the probabilistic model described in the main text for posing the reconstruction problem, we define $P(\bm{p}|\beta,J)$ as the probability of observing the data $\bm{p}$. Using the actual measured values of $\bm{p}$, $P(\bm{p}|\beta,J)$ defines the likelihood of the parameters $\beta$ and $J$,
\begin{widetext}
\begin{equation}
P(\bm{p}|\beta,J)=\frac{1}{Z(\beta,J)}\int d^N\bm{x} e^{-\frac{\beta}{2}(\bm{A}\bm{x}-\bm{p})^T(\bm{A}\bm{x}-\bm{p})-\frac{J}{2}\bm{x}^T\bm{L}\bm{x}}\prod_{i=1}^N\psi_i(x_i)\, ,
\end{equation}
\end{widetext}
with $Z(\beta,J)$ the normalization factor of $P(\bm{p}|\beta,J)$, so that $\int d^M\bm{p}P(\bm{p}|\beta,J)=1$. Hence, we would  like to find the values of $\beta$ and $J$ such that the likelihood above is maximized. However, due to the functional dependence on the parameters, a direct maximization procedure is rather impractical. The EM algorithm provides an alternative to iteratively estimate the optimal value of $\beta$ and $J$. We can justify such an iteration process if we first introduce the EM free energy functional
\begin{equation}
F_{\text{EM}} = - \log P(\bm{p}|\beta,J)\, ,
\end{equation}
and we then look for the pair $(\beta^\star,J^\star)$ such that the  likelihood is maximized.  Seeking indeed that $\frac{\partial F_{\text{EM}}}{\partial \beta}=\frac{\partial F_{\text{EM}}}{\partial J}=0$, we  obtain
\beeq{
	\beta^\star&=\frac{M}{\bracket{(\bm{A}\bm{x}-\bm{p})^T(\bm{A}\bm{x}-\bm{p})}_{\star}}\,,\\
	 J^\star&=\frac{N}{\bracket{\bm{x}^T\bm{L}\bm{x}}_{\star}}\,,
	\label{eq:para}
}
with
\begin{widetext}
\beeq{
	\bracket{(\cdots)}_{\star}&=\frac{1}{Z(\beta,J)}\int d^N\bm{x} (\cdots)e^{-\frac{\beta^\star}{2}(\bm{A}\bm{x}-\bm{p})^T(\bm{A}\bm{x}-\bm{p})-\frac{J^\star}{2}\bm{x}^T\bm{L}\bm{x}}  \prod_{i=1}^N\psi_i(x_i)\,.
}
\end{widetext}
This provides a closed set of equations for the pair ($\beta^\star$ $J^\star)$, which is solved by  the fixed-point  iteration method. Such a procedure will yield the same equations that a direct implementation of EM  would \cite{dempster1977maximum}. On the other hand, as the averages appearing in the formulas \eqref{eq:para} are rather difficult to calculate (as they involve the computation of all the covariances), we estimate them using EP, further assuming that the corresponding distribution is a Dirac delta centered at $\bm{x}=\bracket{\bm{x}}_{Q^{\text{EP}}}$. This finally  results  into:
\beeq{
	\beta^\star&=\frac{M}{(\bm{A}\bracket{\bm{x}}_{Q^{\text{EP}}}-\bm{p})^T(\bm{A}\bracket{\bm{x}}_{Q^{\text{EP}}}-\bm{p})}\,,\\
	 J^\star&=\frac{N}{\bracket{\bm{x}}_{Q^{\text{EP}}}^T\bm{L}\bracket{\bm{x}}_{Q^{\text{EP}}}}\,.
}

\section{Other methods for reconstruction}
In this section we briefly review the other reconstruction algorithms we have compared our results to.
\subsection{Quadratic Programming}

The reconstruction problem, $\bm{A}\bm{x}=\bm{p}$, with the prior information about $\bm{x}$ coming from the Laplacian matrix and in the noiseless scenario, can  be recast  as a constrained quadratic minimization problem:
\begin{equation}
\bm{x}^* = \arg\min_{\mathclap{\substack{\bm{x}:\bm{A}\bm{x}=\bm{p}\\ \bm{x}_{\rm{inf}}\leq \bm{x} \leq \bm{x}_{\rm{sup}}} } }  \quad \bm{x}^T \bm{L} \bm{x} \, .
\end{equation}
Here we have used $\bm{x}_{\rm{inf}}$ and $\bm{x}_{\rm{sup}}$ to denote the lower and upper limits for the pixel values, to mimic the constraint of the interval prior. When dealing with binary images, the pixels of the above solution, whose value are larger than $0.5$ are set to $1$, or to $0$ otherwise.
In the noisy regime, we instead assume to know $\sigma = \beta^{-1/2}$ and we minimize instead:
\begin{equation}
\bm{x}^* = \arg\min_{\mathclap{\substack{\bm{x}_{\rm{inf}}\leq \bm{x} \leq \bm{x}_{\rm{sup}}} } }  \quad J\bm{x}^T \bm{L} \bm{x} + \beta \left( \bm{A}\bm{x} - \bm{p}\right)^T \left( \bm{A}\bm{x} - \bm{p}\right)\,,
\end{equation}
for different values of the parameter $J$. For each trial $J$ we compute the reconstruction error and we keep the smallest one.

\subsection{Total Variation}

As explained in the main text, we can pose the reconstruction problem as an optimization one whose objective function is the $\ell_2$ norm of the error: $||\bm{A}\bm{x}-\bm{p}||_2$. TV is an improvement on this approach, by adding the requirement that the solution also minimizes the $\ell_1$ norm of the image-gradient, $||\nabla_{\rm{img}}\bm{x}||_1$, which is defined as
\begin{equation}
  (\nabla_{\rm{img}} \bm{x})_i = ( x_{i_x}-x_i, x_{i_y}-x_i)\, ,
\end{equation}
where $i_x$ and $i_y$ denote the neighboring pixel  to the right and below $i$, respectively.
Hence, the TV optimization problem reads:
\begin{equation}
\bm{x}^* = \arg\min_{\bm{x}} \ ||\bm{A}\bm{x}-\bm{p}||_2 + \lambda ||\nabla_{\rm{img}}\bm{x}||_1 \, .
\end{equation}
In this last equation, $\lambda$ is a parameter to weight the relevance of the image gradient regularization.
In the case of noisy measurements we repeat the minimization for different values of the parameter $\lambda$ and we report the $E_{2}$ error from the best reconstruction.\\
\\
Since norms are convex functions and the Laplacian matrix is positive semidefinite, the solution to these two optimization problems can be found using convex optimization techniques. For the cases studied in this work, we utilized the \texttt{Convex.jl} and \texttt{Gurobi} optimization packages \cite{convex,gurobi} to find the solution $\bm{x}^*$.

\subsection{Belief Propagation}

The reconstructions using the Belief Propagation algorithm were obtained with the implementation referenced in \cite{Gouillart2013}, which can be found in \cite{bpfortomo}.

\section{Details in the generation of the phantom images and projection matrices}
\label{app:G}

For binary images we used the procedure described in \cite{Gouillart2013} and the script provided in \cite{bpfortomo} to generate a sample of 50 images of size $50\times 50$, whose number of clusters is controlled by an integer parameter $p$ in the following manner: the algorithm generates images within a circle, as shown in the insets of Fig. \ref{fig1} in the main text, and therefore the effective number of pixels to be reconstructed is reduced to 1959. Once the value of $p$ is specified, $p^2$ pixels are chosen randomly as centroids for a Gaussian filter. Once the filter is applied, only the pixels that have a value above the image average value are set to $1$, and the rest of them are set to $0$. On the other hand, when dealing with gray-scale images, the same procedure is used to generate binary clusters, but once they have been constructed, we set the value of all the pixels within one of them to a random integer inside the interval $[105,255]$. This is done for each of the formed clusters.

Finally, the projection matrix $\bm{A}$ is structured tomographically when using the BP algorithm, that is, several parallel rays are projected along a single direction, and this is repeated for angles between $0^\circ$ and $180^\circ$ in regular steps. Whereas $\bm{A}$ was built using single ray projections along random directions for all the other methods. Importantly, for non-binary images the entries of the projection matrix correspond to the length of the ray passing through that pixel, while for binary images its entries are 1 or 0, depending on whether a ray passed or not through the associated pixel.

\bibliography{refs-PNASDraft}

\end{document}